\documentclass[fleqn,usenatbib]{mnras}

\usepackage{newtxtext,newtxmath}
\usepackage{appendix}

\usepackage[T1]{fontenc}
\usepackage{ae,aecompl}
\usepackage{soul}

\usepackage{graphicx}	
\usepackage{amsmath}	
\usepackage{amssymb}	
\usepackage[dvipsnames]{xcolor}
\usepackage{natbib}

\def\equationautorefname~#1\null{equation~(#1)}
\newcommand{\autorefp}[1]{%
  \begingroup%
  \def\equationautorefname~##1\null{equation~##1\null}%
  \autoref{#1}%
  \endgroup%
}

\DeclareMathAlphabet{\mathpzc}{OT1}{pzc}{m}{it}\definecolor{purple}{RGB}{160,32,240}
\newcommand{\rev}[1]{\textcolor{black}{#1}}

\newcommand{\Mh}{M_{\mathrm{h}}}
\newcommand{\Msun}{\,M_{\odot}}

\newcommand{\Mgc}{M_{\mathrm{GC}}}
\newcommand{\Mgcobs}{M_{\mathrm{GC,obs}}}

\newcommand{\Mstar}{M_{\star}}
\newcommand{\ttid}{t_{\mathrm{tid}}}
\newcommand{\tiso}{t_{\mathrm{iso}}}

\newcommand{\feh}{\mathrm{[Fe/H]}}
\newcommand{\Mc}{M_{\mathrm{c}}}
\newcommand{\Mmax}{M_{\mathrm{max}}}
\newcommand{\Mmin}{M_{\mathrm{min}}}
\newcommand{\Mtot}{M_{\mathrm{tot}}}
\newcommand{\Mlim}{M_{\mathrm{lim}}}
\newcommand{\Mu}{M_{\mathrm{u}}}

\title[Cutoff of the CIMF]{Formation of Globular Cluster Systems II: Impact of the Cutoff of the Cluster Initial Mass Function}

\author[Choksi]{
Nick Choksi$^{1,2}$\thanks{E-mail: nchoksi@berkeley.edu} and
Oleg Y. Gnedin$^{2}$
\\
$^{1}$Department of Astronomy, University of California at Berkeley, Berkeley, CA, 94720, USA\\
$^{2}$Department of Astronomy, University of Michigan, Ann Arbor, MI, 48109, USA
}

\date{Released \today}

\pubyear{2018}

\begin{document}
\label{firstpage}
\pagerange{\pageref{firstpage}--\pageref{lastpage}}
\maketitle

\begin{abstract}
Observations of young star clusters reveal that the high-mass end of the cluster initial mass function (CIMF) deviates from a pure power-law and instead truncates exponentially. We investigate the effects of this truncation on the formation of globular cluster (GC) systems by updating our analytic model for cluster formation and evolution, which is based on dark matter halo merger trees coupled to empirical galactic scaling relations, and has been shown in previous work to match a wide array of observational data. The cutoff masses of $\Mc=10^{6.5}\Msun$ or $10^{7}\Msun$ match many scaling relations: between the GC system mass and host halo mass, between the average metallicity of the GC system and host halo mass, and the distribution of cluster masses. This range of $\Mc$ agrees with indirect measurements from extragalactic GC systems. Models with $\Mc<10^{6.5}\Msun$ cannot reproduce the observed GC metallicity and mass distributions in massive galaxies. The slope of the mass-metallicity relation for metal-poor clusters (blue tilt) for all $\Mc$ models is consistent with observations within their errors, when measured using the same method. We introduce an alternative, more robust fitting method, which reveals a trend of increasing tilt slope for lower $\Mc$. In our model the blue tilt arises because the metal-poor clusters form in relatively low-mass galaxies which lack sufficient cold gas to sample the CIMF at highest masses. Massive blue clusters form in progressively more massive galaxies and inherit their higher metallicity. The metal-rich clusters do not exhibit such a tilt because they form in significantly more massive galaxies, which have enough cold gas to fully sample the CIMF. 
\end{abstract}



\section{Introduction}
\label{sec:Intro}

Studies of globular cluster (GC) systems in galaxies over a vast mass range have revealed surprisingly simple scaling relations. Two of the most accessible properties of globular clusters are their metallicities and their masses. The metallicity is often derived from the colour, via an empirically calibrated transformation, while the masses are derived from the luminosity and colour-dependent mass-to-light ratios. In particular, observations have revealed that the total mass of the GC system is a near-constant fraction of the host halo mass \citep[][]{spitler_forbes09, hudson_etal_2014, harris_etal_2015}. The mean and dispersion in metallicity of the GC system gradually increases with the host halo mass \citep{peng_etal06}. Observations also suggest that the metallicity of the most massive ($M \gtrsim 2\times 10^5\Msun$) metal-poor clusters scales weakly with cluster mass \citep[e.g.,][]{strader_etal09, mieske_etal_2006, cockcroft_etal_2009, harris09a, mieske_etal_2010}. This trend is inferred from the scaling of cluster colour with apparent magnitude, and is therefore often referred to as the ``blue-tilt''. No corresponding trend has been observed for metal-rich clusters. 

In \citet[][hereafter CGL18]{choksi_etal_2018}, we presented an analytic model for the formation of GC systems that matches these trends, based on the earlier models of \cite{muratov_gnedin10} and \cite{li_gnedin14}. The model forms GCs in periods of rapid accretion onto the host dark matter halo. GCs are drawn from a cluster initial mass function (CIMF), and the properties of each cluster are set based on the properties of the host galaxy, which are in turn set using empirically motivated galactic scaling relations. This model successfully reproduces a wide variety of the observed properties of GC systems, including the combined GC mass-halo mass relation, the scaling of the mean metallicity of the GC system, the blue-tilt, and the age-metallicity relation. 

Following previous versions of our model, the CGL18 model adopted a power-law (PL) CIMF with an index $\beta = 2$. This simple functional form was motivated by observations of the CIMF of young massive clusters in nearby star-forming galaxies \citep{zhang_fall99, lada_lada03, portegies_zwart_etal10}. These young clusters have masses and sizes consistent with the properties of objects which could evolve into GCs after a few Gyr of dynamical and stellar evolution. However, detailed modeling of the CIMF of young clusters reveals deviations from pure power-law (PL) behaviour at high masses \citep{gieles_etal_2006a, larsen2009, bastian_2008, adamo_etal_2015, johnson_etal_2017, messa_etal_2018}. Thus, the entire CIMF of young clusters is best described by a Schechter function, $dN/dM \propto M^{-\beta} e^{-M/\Mc}$, where $\Mc$ is the characteristic truncation mass \citep{schechter_1976}.

Another line of evidence supporting the exponential truncation comes from the \textit{present-day} mass function of old GCs, which shows a roughly log-normal distribution with a near-universal peak at $2 \times 10^5\Msun$. Several authors \citep[e.g.,][]{jordan_etal_2007} have shown that this mass function is also well described by an ``evolved" Schechter function, of the form
\begin{equation}
  dN/dM \propto (M + \Delta M)^{-\beta} \exp\left(-\frac{M+\Delta M}{\Mc}\right),
  \label{eq:evolved_mf}
\end{equation}
where $\Delta M$ is the average mass lost by GCs between formation and $z=0$.

\rev{Many works have investigated the physical origin of the maximum mass scales of star forming clumps and stellar clusters \citep[e.g.,][]{dekel_etal_2009, kruijssen_2014, adamo_etal_2015}. These studies have generally suggested that the maximum mass is set by the Toomre mass, corresponding to the maximum mass of a gravitationally unstable clump in a rotationally supported disc \citep{toomre_1964}. Recently, \cite{reina-campos_kruijssen_2017} further noted that feedback from young stars plays an important role in setting the maximum cluster mass by disrupting the cluster before the collapse of a Toomre-unstable region is complete.} 

\rev{\cite{li_etal_sim1, li_etal_sim2} performed high-resolution cosmological simulations in which all star formation is implemented as occurring in star clusters of various masses. Clusters grow via accretion from the local interstellar medium and their growth is terminated self-consistently by on their own feedback. The Schechter-like CIMF with power-law slope of $\beta \approx 2$ and an $\Mc$ that scales with the star formation rate is robustly produced in these simulations. They further show that the maximum cluster mass is sensitive to the star formation efficiency per free fall time $\epsilon_{\rm ff}$, which in turn indirectly sets the strength of stellar feedback in the cluster. \cite{meng_etal_2019} showed that the Toomre analysis can surprisingly accurately predict the unstable regions of the interstellar medium in these simulations, despite the presence of strong turbulent flows. However, they find that the Toomre mass is very large, typically above $10^9\Msun$ at high redshift $z>1.5$, which may be too high to influence the maximum mass of individual clusters formed in these simulations.}

In this work, we update our model to include the exponential truncation of the CIMF and assess the impact of this modification for predictions of GC scaling relations. We begin in \autoref{sec:methodology} with a brief overview of the model. Then we describe our method for incorporating a Schechter function CIMF in \autoref{sec:sampling}. In \autoref{sec:results} we discuss how this change affects the overall agreement between observed and model mass and metallicity distributions. In \autoref{sec:scalings} we present the model predictions for scaling relations of GC system properties using the modified CIMF. In \autoref{sec:bluetilt} we show how the modified CIMF affects the strength of the blue-tilt that arises naturally arises in our model and then we investigate the dependence of the blue-tilt on galaxy assembly histories. \autoref{sec:discussion} discusses the implications of our results and \autoref{sec:summary} summarizes our main conclusion. 

\section{Methodology}
\label{sec:methodology}

Our model predicts GC formation and disruption across the whole of cosmic history. Below we list all the equations required to calculate it, and introduce the cutoff of the CIMF. More details and justification for the choice of equations are provided in CGL18. The two adjustable model parameters\footnote{While the current form of our model has only two adjustable parameters, $p_2$ and $p_3$, we preserve the notation for these parameters for consistency with the past published iterations of our model.} ($p_2,p_3$) are fixed using the comparison with a wide sample of observed GC systems.

\subsection{Summary of cluster formation model}

Cluster formation is triggered when the accretion rate onto a dark matter halo between two consecutive outputs of our adopted dark matter simulation exceeds an adjustable threshold value $p_3$. For a halo of mass $M_{\rm h,2}$ at time $t_2$, and its progenitor of mass $M_{\rm h,1}$ at time $t_1$, we compute the merger ratio, $R_m$, as:
\begin{equation}
  R_{\rm m} \equiv \frac{M_{\rm h,2} - M_{\rm h,1}}{t_2 - t_1} \frac{1}{M_{\rm h,1}},
  \label{eqn:rm}
\end{equation}
and trigger cluster formation at time $t_2$ if $R_{\rm m} > p_3$. \rev{In this work, as in CGL18, we use the properties of dark matter halos from the collisionless run of the \textit{Illustris} cosmological simulation \citep{vogelsberger_etal_2014, nelson_etal_2015}. We note that in  \cite{li_gnedin14}, our cluster formation model was applied to halo merger trees extracted from the \textit{Millenium-II} collisionless simulation, and that we have also tested the model on the EAGLE simulation. In all cases, the results are not sensitive to the adopted simulation.}

Once cluster formation is triggered, we form a population of clusters characterized by total mass $\Mtot$\footnote{This notation differs from that used in CGL18, in which we referred to it as $\Mgc$. To avoid confusion with other quantities, we switch to the label $\Mtot$ in this work, reserving $\Mgc$ for the total mass in GCs at $z=0$ in a galaxy.}, based on the hydrodynamic cosmological simulations of \cite{kravtsov_gnedin05}:
\begin{equation}
  \Mtot = 1.8\times 10^{-4}\, p_2 \, M_g,
  \label{eqn:mgc}
\end{equation}
where $p_2$ is the second adjustable model parameter and $M_g$ is the cold gas mass in the host galaxy. \rev{The purpose of the parameter $p_2$ is to normalize the formation rate of a cluster population in a given episode. It absorbs many factors relevant to cluster formation: the fraction of cold gas in the star-forming phase at that epoch, the efficiency of conversion of that gas into stars, the fraction of new stars in clusters above our adopted minimum mass of $10^5 M_{\odot}$, and the variation of all these factors with the galactic environment. We do not attempt to model all these factors in detail, and instead treat $p_2$ as an adjustable parameter. Typical values of $p_2 \sim 10$ (presented in \autoref{tab:models}) imply that, over a given merger event, $\sim 2 \times 10^{-3}$ of a galaxy's cold gas mass is converted into GCs. This fraction is broadly consistent with numerical simulations of galaxy and cluster formation by \cite{li_etal_sim1}. In these simulations the ratio of the total bound mass of clusters with $M > 10^{5} M_{\odot}$ formed within intervals of 100 Myr to the galaxy gas mass varies by an order of magnitude, but the average value for the epochs satisfying our criterion \ref{eqn:rm} is $M_{\mathrm{tot}}/M_{g} \sim (1-2) \times 10^{-3}$. It is therefore reasonable to fit the average normalization of the cluster formation rate within this range, and even a wider range given the expected variation of the rate over time.}

The cold gas fraction is parameterized as a \rev{function of the stellar mass $\Mstar$ and redshift $z$} as: 
\begin{equation}
  \eta(\Mstar,z) = 0.35 \times 3^{2.7} \, \left(\frac{\Mstar}{10^9\Msun}\right)^{-n_m(\Mstar)} \ \left(\frac{1+z}{3}\right)^{n_z(z)}.
  \label{eqn:fg}
\end{equation}
The redshift and stellar mass scalings, $n_z$ and $n_m$ respectively, are given by:
\begin{align}
  n_z &= 1.4 \;\mathrm{for}\; z > 2, \;\mathrm{and}\; n_z = 2.7 \;\mathrm{for}\; z < 2, \nonumber \\
    n_m &= 0.33 \;\mathrm{for}\; \Mstar > 10^{9}\Msun, \;\mathrm{and}\; n_m = 0.19 \;\mathrm{for}\; \Mstar < 10^{9}\Msun.
  \nonumber
\end{align}
The stellar mass is increased self-consistently using a modified version of the stellar mass-halo mass relation derived from the abundance matching results of \cite{behroozi_etal_2013_main}. 

We then draw individual clusters from the cluster initial mass function, as described in detail in the following section. Each cluster is assigned the average metallicity of its host galaxy at formation, which is set by an observed galaxy stellar mass-metallicity relation: 
\begin{equation}
  \feh =  \log_{10}\left[\left(\frac{\Mstar}{10^{10.5}\Msun}\right)^{0.35} (1+z)^{-0.9}\right].
  \label{eq:mmr}
\end{equation}

\subsection{Monte Carlo sampling of the Schechter function}
\label{sec:sampling} 

For a given formation event, with a combined mass $\Mtot$ to be distributed into individual clusters, we draw clusters from a mass function of the form:
\begin{equation}
  \frac{dN}{dM} = M_0 M^{-\beta} e^{-M/\Mc},
  \label{eqn:cimf}
\end{equation}
where $\beta$ is the index of the power-law, $M_0$ is an overall normalization factor, and $\Mc$ is the truncation mass. As in CGL18, we adopt a constant slope $\beta = 2$. 

Our procedure for drawing clusters is based upon the ``optimal sampling'' method of \cite{schulz_etal_2015}. We begin by drawing the most massive cluster, of mass $\Mmax$. The value of $\Mmax$ is obtained from imposed constraints. The first constraint is that the integral mass equals $\Mtot$:
\begin{equation}
  \Mtot = \int_{\Mmin}^{\Mmax} M\frac{dN}{dM} dM, 
  \label{eqn:constraint1}
\end{equation}
where $\Mmin = 10^5 \Msun$ is the minimum mass of clusters that can form in our model. Clusters with initial masses below $10^5 \Msun$ are expected to be disrupted in $\lesssim 10$~Gyr by the external tidal field. The second constraint is derived from assuming there is only one cluster of mass $\Mmax$:
\begin{equation}
  1 = \int_{\Mmax}^{\infty} \frac{dN}{dM} dM.
  \label{eqn:constraint2}
\end{equation}
Combining both constraints yields $\Mtot$ as a function of $\Mmax$:
\begin{equation}
  \Mtot = \frac{\Gamma(2-\beta, \Mmin/\Mc) - \Gamma(2-\beta, \Mmax/\Mc)}{\Gamma(1-\beta, \Mmax/\Mc)} \, \Mc,
  \label{eqn:constraint3}
\end{equation}
\rev{where $\Gamma(s,x)$ is the upper incomplete gamma function.} We solve this equation numerically for $\Mmax$. After drawing the most massive cluster we calculate the cumulative distribution, $r = N(<M)/N(<\Mmax)$ and invert it numerically. We then draw clusters by sampling the cumulative distribution for $0 \leq r \leq 1$ until the total mass in clusters reaches $\Mtot$. \rev{In Appendix \ref{sec:appendix_sampling} we discuss the effects of adopting alternate sampling methods.}

The realistic value of $\Mc$ may depend on local properties of the ISM such as pressure and density \citep[e.g.,][]{kruijssen_2014}. However, our model contains no spatial information and therefore a detailed calculation of the local value of $\Mc$ is beyond the scope of this work and would only significantly complicate the model. Instead, we choose a constant value of $\Mc$ throughout the calculation and analyze the impact on the model for a few different values of $\Mc$.

\citet{jordan_etal_2007} fit the evolved Schechter function (\autorefp{eq:evolved_mf}) to GCs in the Virgo Cluster Survey (VCS) by assuming a constant mass offset $\Delta$ for all clusters in a given galaxy. Their results were updated by \citet{johnson_etal_2017}, who included stellar evolution mass loss and revised stellar mass-to-light ratios. Both of these changes affect the fractional mass loss of each cluster and therefore rescale the expected initial cluster mass by a constant factor ($\approx 2.1$). This in turn results in an increase of the fitted cutoff mass by the same factor. For host galaxies with the stellar mass $10^{9}-10^{12}\Msun$, \citet{johnson_etal_2017} find values of $\Mc$ ranging from $10^6-10^7\Msun$ and a weak scaling with galaxy mass. These inferred values are still expected to underestimate the true value of $\Mc$, because the assumption of a constant mass offset for all GCs is inconsistent with the nonlinear scaling of the disruption time with cluster mass (see \autorefp{eqn:ttid} below). The most massive clusters, which determine the fitted value of $\Mc$, experience a larger $\Delta M$ than low-mass clusters. This effect should push $\Mc$ even higher.

Our model sample covers the range of dark matter halo mass from $10^{11}-10^{14.5}\Msun$, which maps to a range of median stellar masses similar to the observed galaxy sample used in \cite{johnson_etal_2017}. Motivated by these results, we test constant values of $\Mc = 10^{6}, 10^{6.5}, 10^{7}, 10^{7.5}\Msun$. We find that lower values of $\Mc < 10^{6}\Msun$ severely truncate the formation of massive clusters that should form in giant galaxies and therefore cannot reproduce the cluster mass functions. Higher values than $10^{7.5}\Msun$ give indistinguishable results to the PL model. The most appropriate value of $\Mc$ to match the overall observed distribution appears to be $\Mc \approx 10^{6.5}-10^7\Msun$.

\subsection{Cluster disruption}
\label{sec:disruption}

CGL18 adopted a modified version of the analytic cluster disruption prescription used in \cite{gnedin_etal14}, which accounts for dynamical disruption in the presence of both a strong and weak tidal field:
$$
  \frac{dM}{dt} = -\frac{M}{\mathrm{min}\left(\tiso, \ttid\right)},
$$
where $\tiso$ and $\ttid$ are the disruption timescales in the isolated, weak tidal field limit and strong tidal field limit, respectively. However, this prescription did not take into account the expansion of the cluster as two-body relaxation progresses, and thus overestimated the importance of disruption in isolation. To avoid determining a new prescription for $\tiso$, which is beyond the scope of this work, we simplify the disruption prescription to include only disruption in the strong tidal field limit. This change is reasonable, because in the CGL18 model $\ttid < \tiso$ for $M \geq 5\times 10^3\Msun$, so  disruption in isolation only affected the lowest mass clusters.  The final prescription is then: 
\begin{equation}
	\frac{dM}{dt} = -\frac{M}{\ttid},
    \label{eqn:dmdt}
\end{equation}
where $\ttid$ is:
\begin{equation}
  \ttid(t) \approx 5\,\mathrm{Gyr}\, \left(\frac{M(t)}{2 \times 10^{5}\Msun}\right)^{2/3} \left(\frac{P}{0.5}\right),
  \label{eqn:ttid}
\end{equation}
and $P$ is a normalized period of rotation around the galactic center\rev{, defined in \cite{gnedin_etal14}}. As in CGL18, we adopt a constant value of $P=0.5$.   Integrating \autoref{eqn:dmdt} gives the mass evolution from dynamical disruption:
\begin{equation}
  M'(t) = M_0\left[ 1 - \frac{2}{3}\frac{t}{t_{\mathrm{tid}}(t=0)} \right]^{3/2}.
\end{equation}
We count time $t$ from the formation of each cluster individually. In addition to dynamical disruption, we include a time-dependent mass-loss rate due to stellar evolution, $\nu_{\rm se}$, as calculated by \cite{prieto_gnedin08}, and assume it occurs much faster than the dynamical disruption. The combined cluster mass evolution is then:
\begin{align}
  M(t) = M'(t) \left[ 1 - \int_0^t \nu_{\rm se}(t')dt' \right].
\end{align}

\begin{table}
\centering
\begin{tabular}{llccccr}
\hline\\[-2mm]
Model & $\Mc\,(M_{\odot})$ & $p_2$ & $p_3\,(\mathrm{Gyr}^{-1})$ & $G_{Z}$ & $G_{M}$ & ${\cal M}$ \\[1mm]
\hline\\[-2mm]
S6   & $10^{6}$   & 21 & 0.75 & 0.45   & 0.22 & 12.8 \\
S6.5 & $10^{6.5}$ & 13.5 & 0.70 & 0.51 & 0.40 & 9.1 \\
S7   & $10^{7}$   & 8.8  & 0.58 & 0.54 & 0.57 & 7.9 \\
S7.5 & $10^{7.5}$ & 7.15  & 0.50 & 0.57 & 0.61 & 7.6 \\
PL   & $\infty$   & 6.75 & 0.50 & 0.58 & 0.67 & 7.3 \\
\hline
\end{tabular}
\caption{\small Best fit parameters for different cutoff masses of the Schechter function, the associated metallicity and mass ``goodness'' values (see \autoref{sec:optimization}), and the merit function value.}
  \label{tab:models}
\end{table}

\subsection{Parameter optimization}
\label{sec:optimization}

For each value of $\Mc$, we search for new best values of $p_2$ and $p_3$ using the same method as in CGL18. We minimize the ``merit function'', $\cal M$: 
$$
  {\cal M} \equiv
  \frac{1}{N_h} \sum_h \left(\frac{\Mgc(z=0)}{\Mgcobs(\Mh)} - 1 \right)^2
  + \frac{1}{N_h}\sum_{h}  \left( \frac{0.58}{\sigma_{Z, h} } \right)^2
  + \frac{1}{G_{M}} + \frac{2}{G_{Z}}.
$$
The first term in the merit function gives the reduced $\chi^2$ of the total GC system mass-halo mass relation. The second term weights the dispersion of the metallicity distributions of model halos against the mean observed value of 0.58~dex. The final two terms weight the ``goodness'' of the metallicity and mass distributions ($G_Z$ and $G_M$, respectively). In brief, they are defined as the fraction of observed-model GC metallicity or mass distribution pairs that have an acceptable Kolmogorov-Smirnov test probability, $p_{KS} > 1\%$, of being drawn from the same underlying distribution. To make a fair comparison between observed and model galaxies, we calculate the median halo mass corresponding to the stellar mass of each host galaxy, and match the distribution of each observed galaxy against all model halos at $z=0$ within $\pm$0.3~dex in mass. The observational data used are a compilation from the VCS and the HST-BCG survey \citep{cote_etal06, peng_etal06, harris_etal14_bcg1, harris_etal16_bcg2, harris_etal17_bcg3}.

Throughout, we refer to our power-law CIMF model as PL, and Schechter function models as S$N$, where $N$ is the adopted value of $\log_{10}\Mc/\Msun$. The best-fit parameter values and the associated values of $\cal M$ are given in \autoref{tab:models}.

\section{Results}
\label{sec:results}

We find that decreasing $\Mc$ worsens both goodness parameters and the residual value of the merit function, after optimizing the parameters $p_2$ and $p_3$. The lowest cutoff mass we consider, $\Mc = 10^{6}\Msun$, cannot adequately match the observed mass or metallicity distributions. The merit function value is 75\% larger and the GC mass function is consistent with the corresponding observational analogs in only 22\% of the cases. The match of the metallicity distribution is also below 50\%, which we consider unacceptable.

However, the choice of $\Mc = 10^{6.5}\Msun$ is acceptable. Relative to the PL model, the goodness parameters $G_Z$ and $G_M$ are reduced only by 0.07 and 0.27, respectively. This is a reasonable match to the data, given the simplicity of the model. The choice of $\Mc = 10^{7}\Msun$ works even better, giving the goodness of both mass and metallicity functions above 50\%. Since $\Mc$ delineates the maximum of the observed range, we consider it as the highest viable value to investigate in detail, along with the preferred $10^{6.5}\Msun$. \rev{Higher values of the cutoff mass produce results close to the original PL model, but are disfavored by modeling of the present-day GC mass function \citep{jordan_etal_2007, johnson_etal_2017}.}

\begin{figure}
\includegraphics[width=\columnwidth]{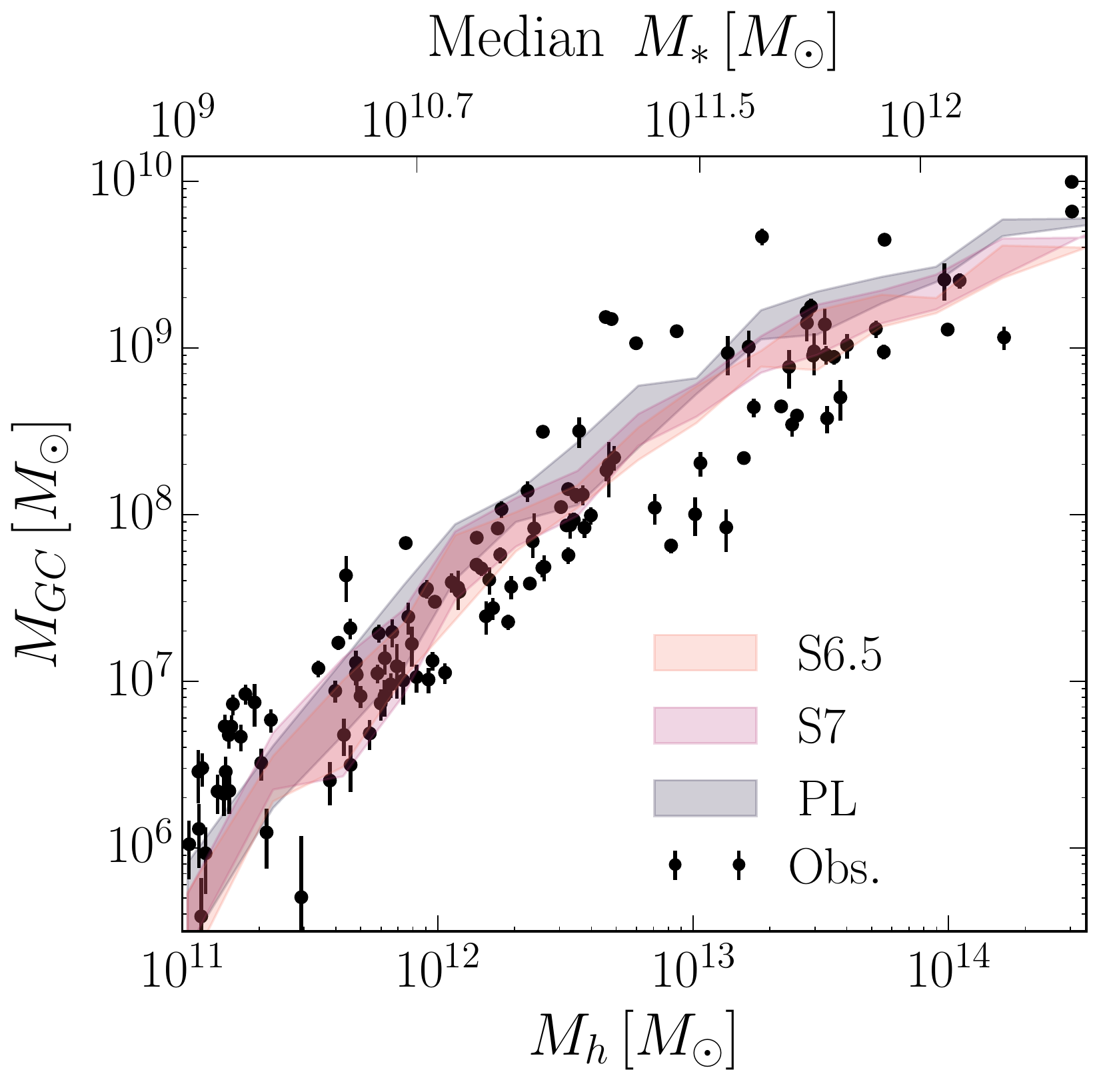}
\vspace{-5mm}
\caption{Combined mass of all GCs as a function of host halo mass, both at $z=0$. In addition to the original PL model from CGL18, we show two models with the values of cutoff mass closest to the distribution inferred by \citet{johnson_etal_2017}: $\Mc = 10^{6.5}$ and $10^{7}\Msun$. Observed halo masses are estimated from weak lensing by \protect \cite{hudson_etal_2014} and \protect\cite{harris_etal_2015}.}
  \label{fig:m_sch}
\end{figure}

\subsection{Scaling relations of globular cluster systems}
\label{sec:scalings}

In \autoref{fig:m_sch} we show the relation between the total mass of the GC system, $\Mgc$, and its host galaxy halo mass, $\Mh$, at $z=0$. The $\Mgc-\Mh$ relation remains robust despite the introduction of the cutoff mass. For most large halos with $\Mh > 10^{11.5}\Msun$, the models with smaller $\Mc$ match the data even better than the PL model. As a simple estimate of a match between a model and the observed data, we compute the \textsc{rms} deviation of the data from the median trend in each model (in log mass) and divide it by the standard deviation of the data, $\sigma$, in 0.5~dex bins of halo mass. All bins except the lowest-mass bin have \textsc{rms}$/\sigma < 1$, and lower values for lower $\Mc$. Only at $10^{11}\Msun < \Mh < 10^{11.5}\Msun$ does the deviation increase with decreasing $\Mc$ and reach \textsc{rms}$/\sigma \approx 2$. 

The best-fit value of the normalization of the formation rate $p_2$ increases with decreasing $\Mc$, and therefore a larger total mass is initially formed in clusters ($p_3$ changes only slightly). However, at $z=0$ the total mass of the GC system $\Mgc$ in the S6.5 (S7) model is \textit{lower} than in the PL case by 0.15 (0.1) dex on average. Because the lower-$\Mc$ models preferentially form lower-mass clusters, they lose these clusters faster (see \autorefp{eqn:ttid}): 3 (1.8) times as many clusters are disrupted by $z=0$ in the S6.5 (S7) model as in the PL model.

\begin{figure}
\includegraphics[width=\columnwidth]{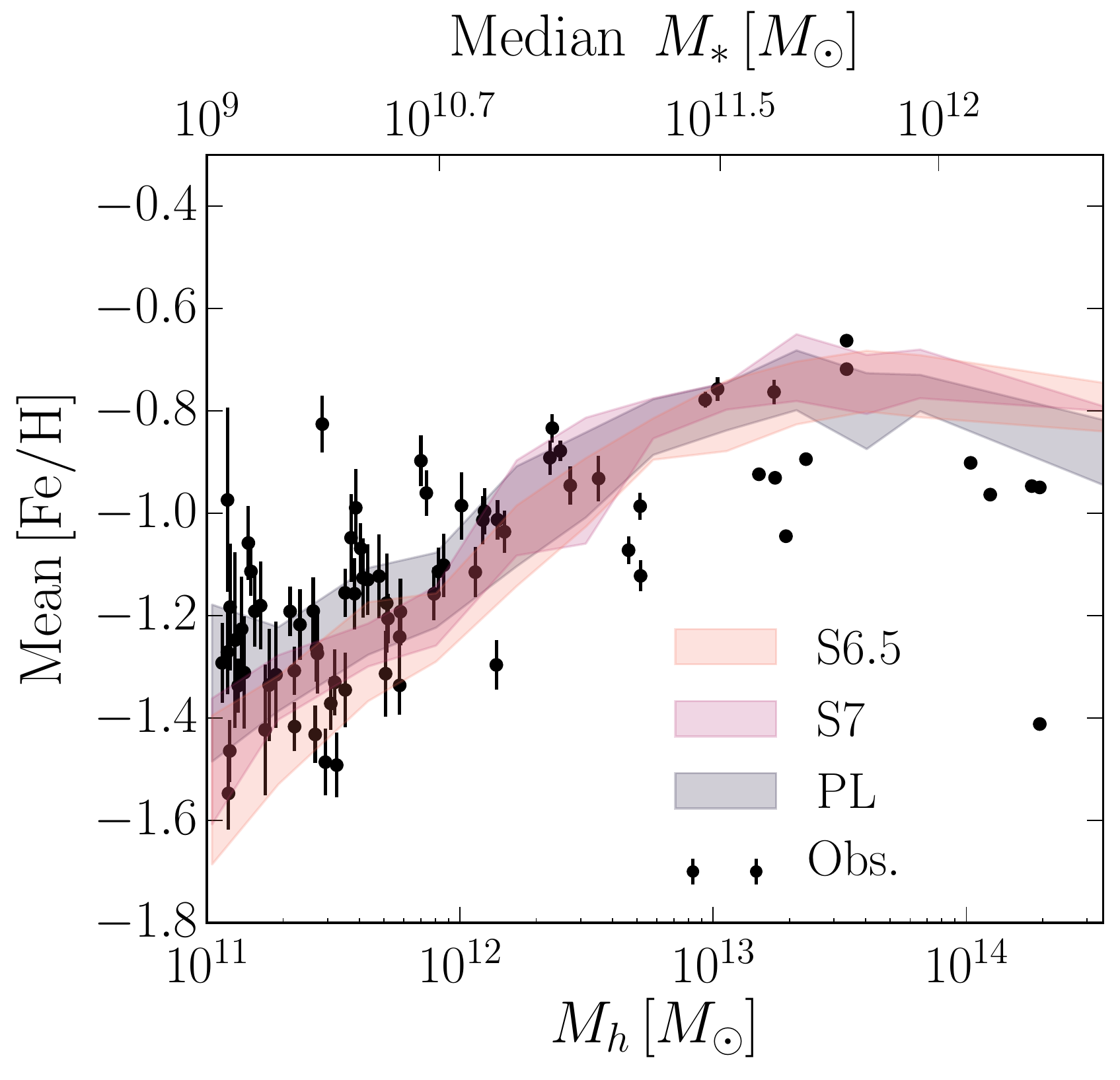}
\vspace{-5mm}
\caption{Mean metallicity of GC system as a function of the host halo mass at $z=0$. Data points are a compilation from the Virgo Cluster Survey and HST-BCG, scaled to the metallicity calibration of the VCS. Halo masses are calculated using the stellar mass-halo mass relation of \protect\citet{kravtsov_etal_2014}, and stellar masses are computed using the color-dependent mass-to-light ratios of \protect\citet{bell_etal03}.}
  \label{fig:mean_sch}
\end{figure}

\autoref{fig:mean_sch} shows the scaling of the mean metallicity of the GC system with host halo mass for the different $\Mc$ models. The rms deviation is larger than for the $\Mgc-\Mh$ relation, \textsc{rms}$/\sigma \lesssim 1.5$, and generally increases with decreasing $\Mc$. In the most massive halo bin, $\Mh > 10^{14}\Msun$, the ratio reaches \textsc{rms}$/\sigma \approx 4$. We plan to investigate this discrepancy, and its relation to the contribution of satellite galaxies, in future work. 

At low halo masses ($\Mh \sim 10^{11} \Msun -10^{11.5} \Msun$), lower $\Mc$ models have systematically lower mean metallicity. \rev{This is because parameter optimization leads to a higher value of $p_2$ for lower $\Mc$ models, producing more blue clusters in small galaxies, thus lowering the mean metallicities.} On the other hand, at higher halo masses, lower $\Mc$ models have \textit{higher} mean metallicities than the PL model. In particular, the observed systems show a break in the scaling of the mean metallicity with halo mass in the most massive hosts, with the mean metallicity instead decreasing slightly, by 0.1~dex. While the qualitative trend of the flattening of the relation at these halo masses is robust, models S6.5 and S7 exhibit a shallower decline in the mean $\feh$ at $\Mh \gtrsim 10^{14}\Msun$. To further understand these results, we examined the distribution of cluster populations $\Mtot$ (\autorefp{eqn:mgc}) that form in qualified \rev{cluster formation} events. When $\Mtot < \Mc$, the cutoff in the mass function is irrelevant: the total mass forming in clusters is too low to sample the high-mass end of the CIMF. When $\Mtot \gtrsim \Mc$, the cutoff becomes important. We find that metal-poor clusters typically form in populations of $\Mtot \sim 10^6-10^{7.5}\Msun$ (interquartile range) and therefore are somewhat affected by the cutoff in our preferred models S6.5 and S7. In contrast, the metal-rich clusters typically form in populations of $\Mtot \sim 10^{6.6}-10^{8.5}\Msun$ because their host galaxies are larger, and therefore our choice of $\Mc$ affects them more significantly. This effect causes more clusters to form in later events, which inherit higher metallicity, as seen in Fig.~\ref{fig:mean_sch}. The mean metallicity of surviving clusters in the most massive halos, $\Mh \gtrsim 10^{13}\Msun$, increases relative to the PL model.

The scaling and normalization of the metallicity dispersion of GC systems, \rev{which we presented and analyzed in CGL18}, is indistinguishable for the different models, and therefore not shown here for brevity.

\begin{figure}
\includegraphics[width=\columnwidth]{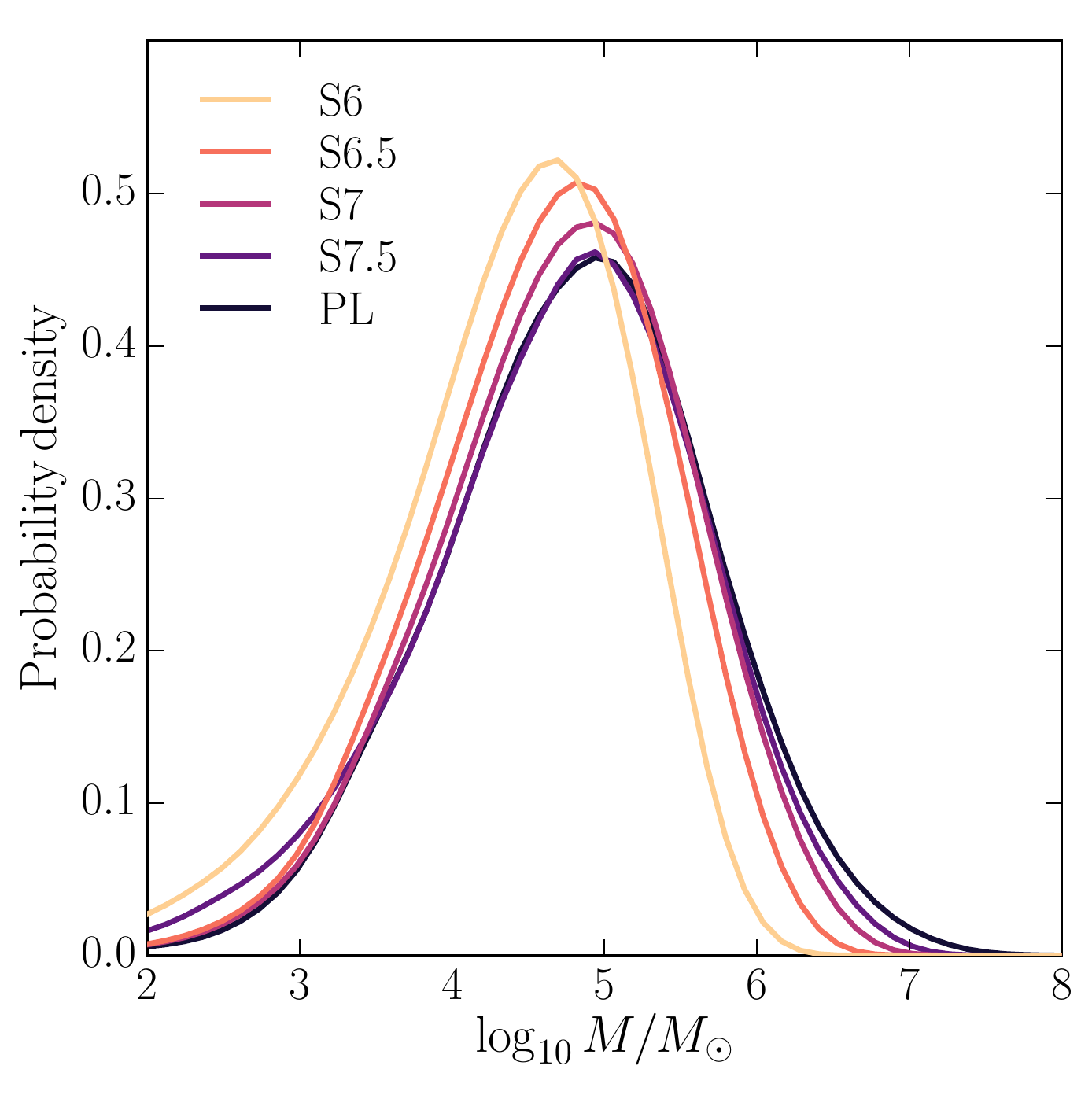}
\vspace{-5mm}
\caption{GC mass function for different models at $z=0$. The distributions have been weighted by the halo mass function so as to be cosmologically representative.}
  \label{fig:mf_stack}
\end{figure}

\begin{figure}
\includegraphics[width=\columnwidth]{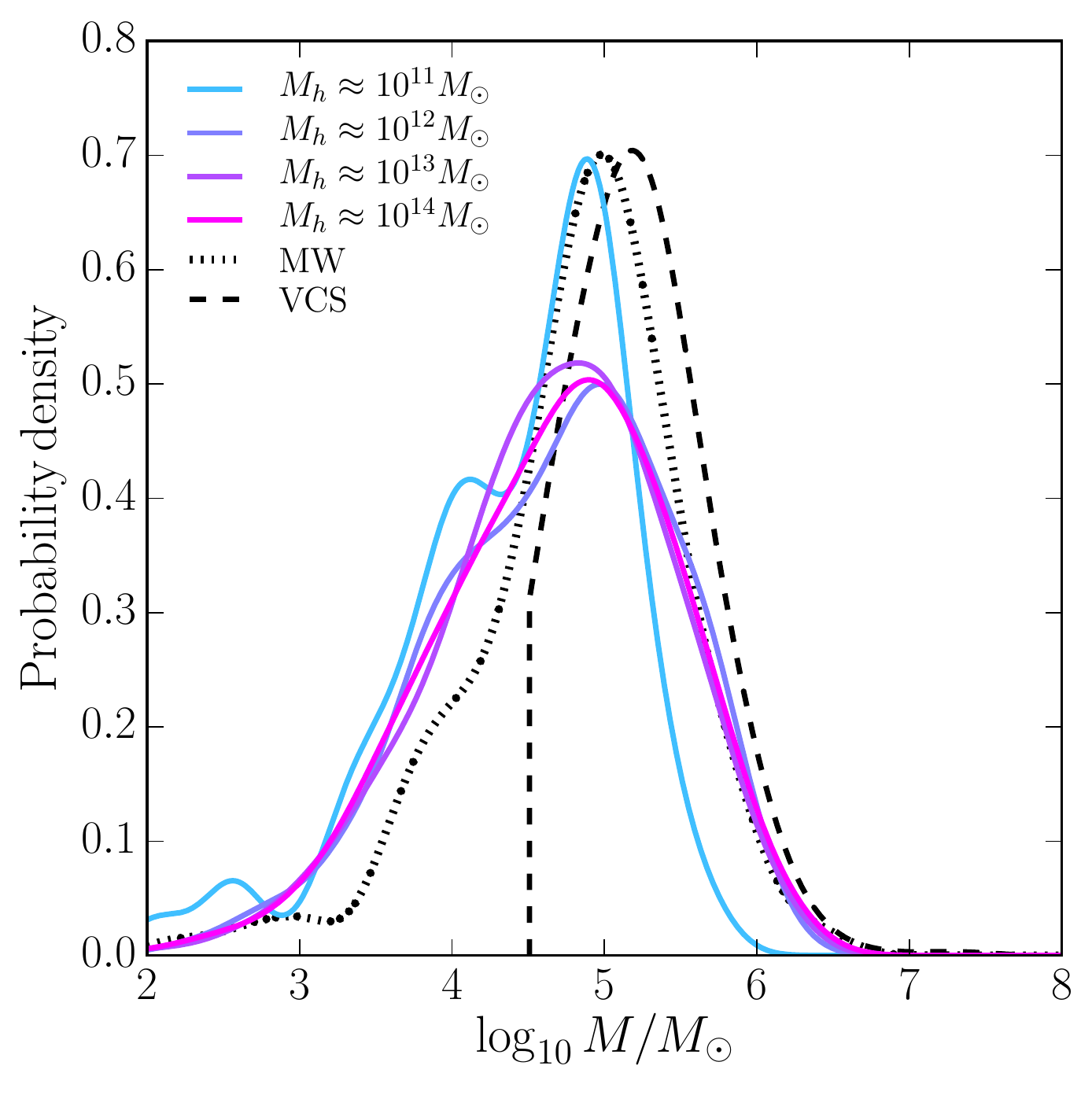}
\vspace{-5mm}
\caption{Examples of the GC mass function for four individual halos at $z=0$ in the $\Mc = 10^{6.5}\Msun$ model, varying by factors of ten in halo mass. The dashed curve shows the mass function of Galactic GCs with updated mass-to-light ratios from \citet{harris_etal_2017}; the dashed curve shows the average mass function of GCs in VCS galaxies (see \autoref{subsec:mf_evolution} for detail).}
  \label{fig:mf_panels}
\end{figure}

\subsection{Evolution of the cluster mass function}
\label{subsec:mf_evolution}

\autoref{fig:mf_stack} compares the total GC mass function (GCMF) at $z=0$ for the different $\Mc$ models. These GCMFs are constructed to represent the total GC sample in a cosmological volume, as would be observed in large-scale surveys. Since the mass function can vary between galaxies of different mass, we weight the contribution of each cluster to the GCMF by the cosmological halo mass function at $z=0$, $n_{\rm h}(\Mh, z=0)$. To do so, we split our sample of model halos into bins of mass of 0.3 dex. Then, we weight each cluster by $n_{\rm h}(\Mh, z=0)$ divided by the number of halos in that mass range that were used in our model. The latter step is needed because to run the model for different mass galaxies we chose a log-linearly spaced subset of halos from the cosmological simulation \textit{Illustris}. The weighting converts that selection to a cosmologically representative sample. The halo number densities were calculated using the analytic halo mass functions of \cite{sheth_tormen99} implemented in the \textsc{colossus} code \citep{diemer_colossus}.

The GCMFs of all models are peaked around $M \approx 10^5\Msun$. Smaller $\Mc$ leads to stronger truncation of the high-mass end, as expected. The mean value and the width of the mass function decrease monotonically with decreasing $\Mc$. The differences between the PL model and $\Mc = 10^6 \Msun$ model mass functions are 0.3~dex in the mean mass and 0.2~dex in the standard deviation.

\autoref{fig:mf_panels} shows several examples of the GCMF for individual galaxies for our preferred choice $\Mc = 10^{6.5}\Msun$. They are all consistent with a universal GCMF. For comparison, we show the Galactic GCMF and a stacked sample of GCs for VCS galaxies. The masses of Galactic clusters were computed by combining their luminosities from the \citet{harris10} catalog and the luminosity-dependent mass-to-light ratios suggested by \citet{harris_etal_2017}. In the VCS, the most luminous galaxies host the majority of the detected GCs in the survey. To obtain an average GCMF, we weight each cluster by the inverse of the number of clusters in its host galaxy. Unlike the case for the Galactic GCs, the VCS sample also suffers from incompleteness. Therefore, we adjust the normalization of the VCS GCMF by the expected fraction of clusters above the detection limit. In CGL18, we estimated this detection limit to be about $10^{4.5}\Msun$. The expected fraction is calculated by integrating the evolved Schechter mass function (\autorefp{eq:evolved_mf}) evaluated for $\Mc = 10^{6.5}\Msun$ and $\Delta M \approx 10^{5.7}\Msun$. Here, $\Delta M$ is the average amount of mass lost by clusters, which is similar in our model and in the fits of \citet{jordan_etal_2007}. Note that we adopt constant values for $\Delta M$ and $\Mc$, while in reality both of these quantities depend on the properties of the host galaxy.

The high-mass galaxies match the Galactic GCMF well, except for the somewhat wider distribution extending to low cluster masses. The only significant deviation is in the smallest halo, which lacks clusters above $10^6\Msun$. However, the peak mass is very consistent at $\approx 10^{5}\Msun$ for all the galaxies. 

The GCMF of the VCS galaxies is slightly offset from the Galactic GCMF. This illustrates the modest galaxy-to-galaxy variation of the mass function and the importance of considering all available datasets when comparing model predictions with observations. Our S6.5 model does not produce as narrow a mass distribution as in the VCS and misses the most massive clusters. However, model S7 (not shown for brevity) can match the VCS GCMF. It may be that $\Mc \approx 10^{7}\Msun$ is required to describe the GC systems of the early-type galaxies in Virgo cluster, while $\Mc \approx 10^{6.5}\Msun$ is more appropriate for the Milky Way-type galaxies. This is consistent with a general increase of $\Mc$ with host galaxy mass, as suggested by \cite{jordan_etal_2007} and \cite{ johnson_etal_2017}.

\rev{\cite{jordan_etal_2007} and \cite{harris_etal_2014} noted that the typical width of the GCMF scales weakly with the mass of the host galaxy. To investigate this trend in our model, we fit Gaussians to our model GCMFs and checked the scaling of the best-fit standard deviations $\sigma_{\log M}$ with host galaxy $\Mstar$. To ensure fair comparison to observation, we performed the fits only to clusters with $M > 10^{4.5} \Msun$ and adjusted the normalization according to the completeness of the observed sample. Over the range $\Mstar = 10^{10.5} - 10^{12} \Msun$, observations show the average $\sigma_{\log M}$ increases from 0.40 dex to 0.5 dex. Our S6.5 model shows a similar trend, but with a slightly higher normalization: the average $\sigma_{\log M}$ increases from 0.45 to 0.55 dex.}

\rev{The GCMF evolution described above is predicated on the assumed steady tidal mass loss that depends only on cluster mass. In reality, cluster disruption would depend on the environment, including the overall strength of the tidal field and its rapid variation in time (``tidal shocks''). We are unable to include these effects in the present model framework, and therefore cannot conclusively predict how they would influence our conclusions. However, we expect global statistics of many stacked galaxies (as shown in \autoref{fig:mf_stack}) to be robust. For individual galaxies, such as those shown in \autoref{fig:mf_panels}, tidal shocks could be more important. Because on average more massive clusters have higher half-mass density, they are likely to be more resilient to tidal shocks than the less massive clusters. This could potentially reduce the number of surviving low-mass clusters and bring the model mass functions shown in Figure 4 to better agreement with the observations.}

\begin{figure}
\includegraphics[width=\columnwidth]{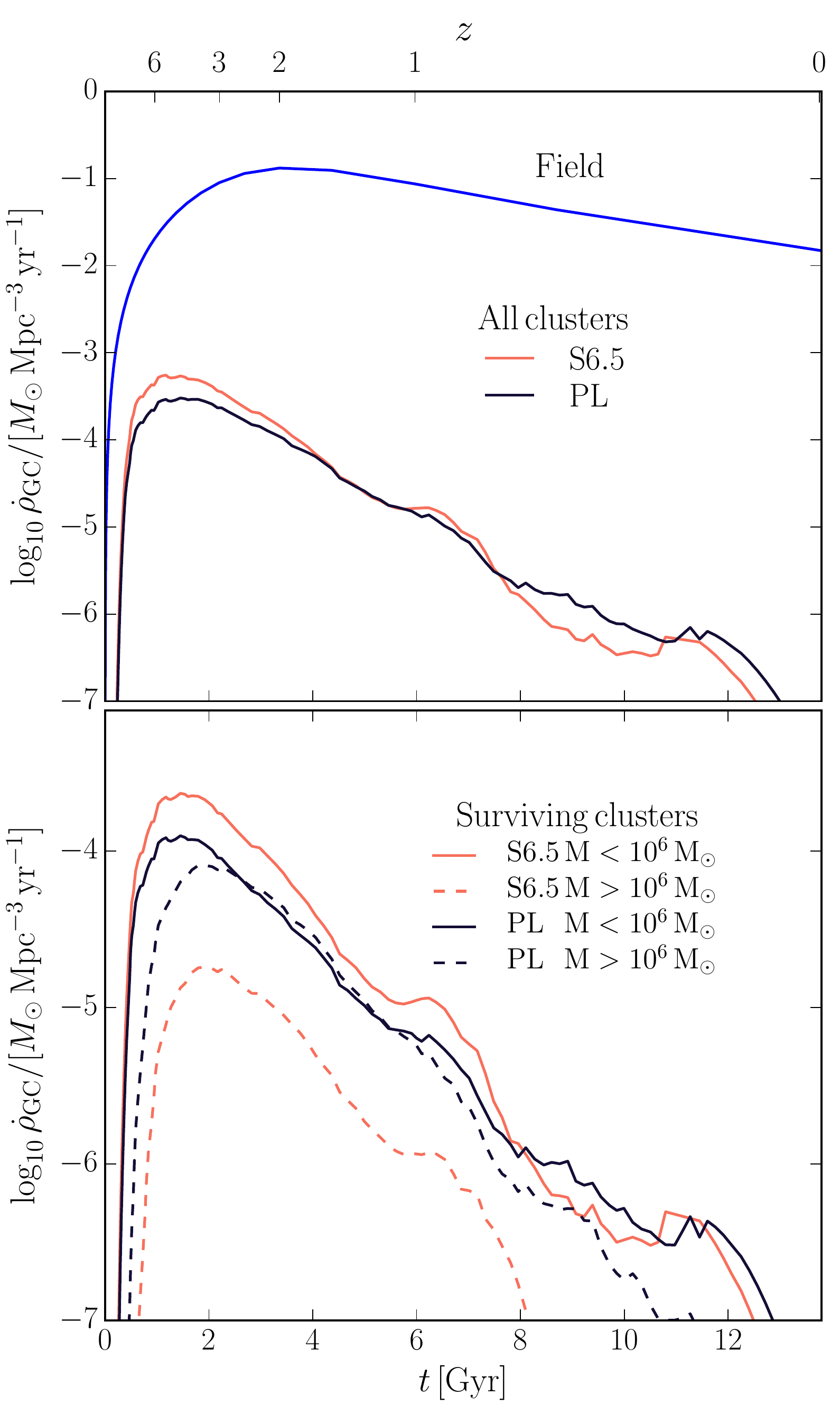}
\vspace{-5mm}
\caption{\textit{Upper:} Cluster formation rate density over cosmic time for the S6.5 and PL models. The blue curve shows the best-fit relation to the observed field cosmic star formation rate density from \protect \cite{madau_dickinson_2014} (extrapolated at $z \gtrsim 8$). \textit{Lower:} Clusters that survive to $z=0$, split into a low and high-mass bin by final mass.}
  \label{fig:gc_sfr}
\end{figure}

\subsection{Formation history}

The top panel of \autoref{fig:gc_sfr} compares the formation rate of GCs with the cosmic star formation history from \citet{madau_dickinson_2014}. In general, the peak of GC formation precedes the peak of the field stellar population by about 2 Gyr, or $\Delta z \approx 2-3$. At its peak epoch, the GC formation rate is of order 1\% of the total star formation rate (SFR). The GC formation rate falls more steeply after the peak than does the field SFR, dropping by four orders of magnitude. Because of the larger normalization $p_2$, the truncated S6.5 model produces a factor of 2 more mass in clusters at high redshift than the PL model.

\citet{el-badry_etal_2019} presented a similar GC formation model to ours, in which the cluster formation efficiency is tied to the gas surface density, which is in turn set by an equilibrium inflow-outflow model. Their model similarly predicts that the GC formation rate peaks earlier than that of the field, with a maximum in the range $z=3-5$. The results of both models show that GCs are unlikely to contribute significantly to the production of ionizing photons before the reionization of cosmic hydrogen at $z\gtrsim6$.

The bottom panel of \autoref{fig:gc_sfr} shows the dramatically different proportions of very massive clusters ($M > 10^6\Msun$) in the S6.5 and PL models. In the PL model, the contributions to the GC formation rate, $\dot{\rho}_{\rm GC}$, from $t=3-8$~Gyr are similar from the low and high-mass clusters. In contrast, in the S6.5 model, massive clusters never make up more than 10\% of $\dot{\rho}_{\rm GC}$.

\begin{figure}
\includegraphics[width=\columnwidth]{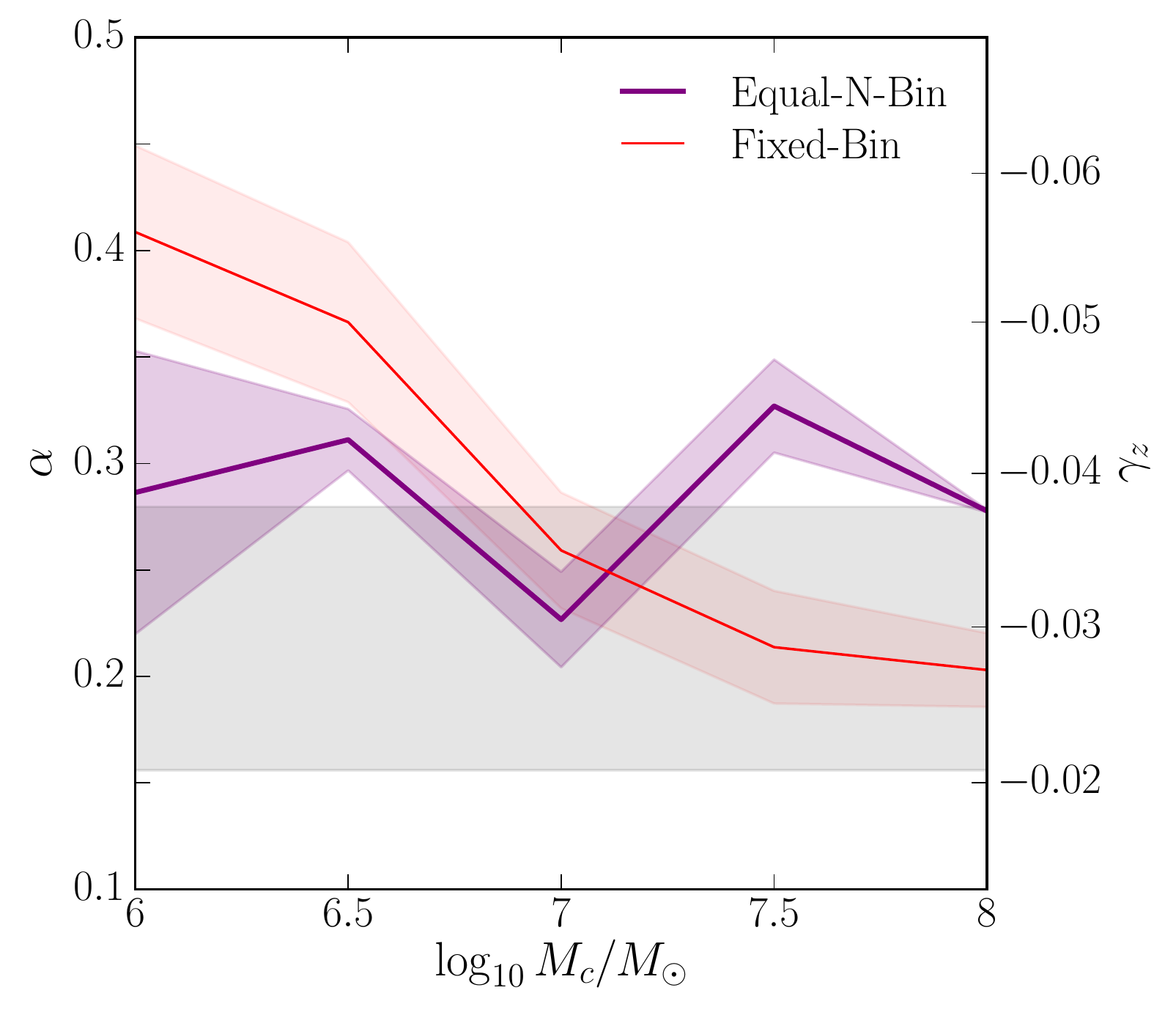}
\vspace{-5mm}
\caption{The blue-tilt slope $\alpha$ for the stacked sample of all clusters in each $\Mc$ model, using the Equal-N-Bin and Fixed-Bin methods (see text for details). Solid lines show the best-fit values and shaded regions show the $1\sigma$ errors. The grey shaded band shows the best-fit slope within its error for the stacked sample of clusters in the Virgo and Fornax Cluster surveys, converted to our metallicity calibration.}
  \label{fig:tilt_slope}
\end{figure}

\section{Steepening of the blue tilt}
\label{sec:bluetilt}

In CGL18, we showed that a correlation between cluster mass and metallicity arises naturally for massive metal-poor (``blue'') clusters. No such trend is found for the metal-rich (``red'') clusters. Describing the relation as
$$ \feh = \alpha \log_{10}M + \mathrm{const}, $$
we found a best value of $\alpha \approx 0.23$ for a stacked sample of all our model clusters with $M \gtrsim 5\times 10^5 \Msun$, consistent with data for observed clusters in the VCS (see Fig.~7 of CGL18).

In our model, this trend arises because the metal-poor clusters form at high redshift ($7 \lesssim z \lesssim 3$) in low-mass halos ($\Mh \lesssim 10^{11} \Msun$). Although the gas fractions of galaxies in this redshift range are high, the \textit{total amount} of cold gas available for cluster formation is relatively low, and as a result very massive clusters cannot typically form in these environments. Therefore, the high-mass end of the CIMF is not fully sampled. Instead, massive blue clusters can form only in galaxies with larger cold gas reservoirs. By the adopted galactic scaling relations, the total cold gas mass scales with the galaxy stellar mass, which in turn scales with the galaxy metallicity (which clusters inherit). In this way, massive blue clusters preferentially form in slightly metal-enhanced environments. Hence, the observed mass-metallicity correlation at $z=0$ is a statistical effect. No correlation is produced for the red clusters, because they form in more massive host halos ($\Mh\sim 10^{11}-10^{13}\Msun$), which have large enough cold gas reservoirs to fully sample the CIMF (see Fig.~8 in CGL18).

The truncation of the CIMF at high masses should affect this model result. In particular, because the probability of forming a massive cluster is exponentially suppressed, the formation of massive blue clusters will be pushed to galaxies with even higher gas masses and metallicities \rev{(recently independently argued by \citealt{usher_etal_2018})}. Thus, a Schechter CIMF should \textit{increase} the strength of the blue tilt -- the value of the parameter $\alpha$. 
\subsection{Quantifying the blue tilt}
How exactly we quantify the blue tilt matters. Our survey of the observational literature showed that two different ways of calculating the blue tilt have been adopted. We find that they produce different results and impact interpretation of trends of the slope $\alpha$. Below we describe these two methods and suggest another method that we think produces more reliable results.

The first method (hereafter ``Equal-N-Bin") follows that used by \citet{mieske_etal_2010} for the combined sample of GCs in the Virgo and Fornax cluster galaxies. We divide clusters above some minimum mass $\Mlim$, usually set by observational completeness, into at most 25 bins of mass with an equal number of clusters in each bin. We take $\Mlim = 10^{5.15}\Msun$ as in \citet{mieske_etal_2010} to allow direct comparison with the observations. Then we fit a sum of two Gaussians to the metallicity distribution of clusters in each bin, representing red and blue clusters. To obtain a reliable fit we need a sufficient number of objects, and therefore we impose a minimum number of clusters per bin: $N_{\rm min, GMM} = 50$. This requirement sets the number of bins as $N_{\rm bins} =  \mathrm{min}\left(N_{\rm tot}/N_{\rm min, GMM}, 25\right)$, where $N_{\rm tot}$ is the total number of clusters in the sample. Using this binning method, in our stack of all model clusters we have between 14,000 and 44,000 clusters per bin, depending on the $\Mc$ model.

For the metallicity distribution fit we use the Gaussian Mixture Modeling (GMM) method described in \citet{muratov_gnedin10}. It gives us the peak locations of the red and blue cluster metallicities, $\mu_{\rm red}$ and $\mu_{\rm blue}$, and the corresponding Gaussian dispersions $\sigma_{\rm red}$ and $\sigma_{\rm blue}$. To ensure robust GMM results, we exclude any bins which do not have the peaks clearly separated. Specifically, if the separation parameter
$$ 
    D  \equiv \frac{|\mu_{\rm blue} - \mu_{\rm red}|}{\left[(\sigma^2_{\rm blue} + \sigma^2_{\rm red})/2\right]^{1/2}} 
$$
for the bin is less than 1.4 then we do not include the bin in the fit. We also exclude bins where the blue peak is ``too red": $\mu_{\rm blue} > -0.8$. These cuts primarily affect the highest mass bins, where the metallicity distributions are nearly unimodal due to merging of the blue and red peaks at high cluster mass.

Given this set of peak metallicities of the metal-poor clusters, we do a linear regression fit between $\mu_{\rm blue}$ and the mean $\log_{10}M$ in the bin:
\begin{equation}
  \mu_{\rm blue} = \alpha \log_{10}\frac{M}{10^6\Msun} + \beta.
  \label{eqn:mu_tilt}
\end{equation}
The pivot at a typical mass of $10^6\Msun$ is chosen to minimize the uncertainty of the intercept $\beta$.

The result of applying the Equal-N-Bin method on the model clusters for several different values of $\Mc$ is shown by the purple curve in \autoref{fig:tilt_slope}. Our model sample includes a stack of clusters from all galaxy halos selected from the \textit{Illustris} cosmological simulation with log-linear spacing in halo mass. Despite our previous argument, this curve shows no correlation between $\alpha$ and $\Mc$. We investigated possible reasons for the lack of a correlation with $\Mc$ and found it to be caused by the variable bin widths used in the Equal-N-Bin method. 

To illustrate the dependence of this result on bin width selection, we test an alternative method of binning clusters for GMM fits: with fixed bin widths of 0.1~dex in mass (hereafter ``Fixed-Bin"). This leads to a median number of clusters per bin between 2,400 and 12,000, depending on the $\Mc$ model. The red curve in Fig.~\ref{fig:tilt_slope} shows the result of using the Fixed-Bin method. Now we see that $\alpha$ decreases monotonically with $\Mc$, consistent with the argument advanced above. In the power-law limit $\Mc \rightarrow \infty$, the slope asymptotes to a value $\alpha \approx 0.20$. Since more massive galaxies are expected to have had larger $\Mc$ at the time of GC formation, fitting the blue-tilt with the Equal-N-Bin method may wash out information about the variation of $\alpha$ with galaxy mass.

\begin{table*}
\centering
\begin{tabular}{llccccccr}
\hline\\[-2mm]
Fit method & Model  & $\alpha$ & $\beta$ & $\gamma_z$ & $\delta_z$ \\[1mm]
  & & 25\% {\bf50\%} 75\% & 25\% {\bf50\%} 75\% & 25\% {\bf50\%} 75\% & 25\% {\bf50\%} 75\% \\[1mm]
\hline\\[-2mm]
Equal-N-Bin& S6.5   & 0.20 {\bf 0.38} 0.56 & -1.18 {\bf -1.10} -1.03 & -0.026 {\bf-0.051} -0.079 & 0.66 {\bf0.69} 0.71 \\
Fixed-Bin  & S6.5  & 0.16 {\bf 0.31} 0.48 & -1.16 {\bf -1.09} -1.04 & -0.022 {\bf-0.042} -0.066 & 0.66 {\bf0.69} 0.71  \\
Single-Split      & S6.5  & 0.19 {\bf 0.24} 0.28 & -1.25 {\bf -1.14} -1.03 & -0.026 {\bf-0.032} -0.037 & 0.63 {\bf0.67} 0.71 \\ \hline
Equal-N-Bin& S7  & 0.21 {\bf 0.28} 0.37 & -1.19 {\bf -1.17} -1.11 & -0.029 {\bf-0.038} -0.050 & 0.65 {\bf0.66} 0.68 \\
Fixed-Bin  & S7   & 0.12 {\bf 0.24} 0.39 & -1.28 {\bf -1.17} -1.10 & -0.015 {\bf-0.033} -0.053 & 0.62 {\bf0.66} 0.68 \\
Single-Split      & S7   & 0.16 {\bf 0.20} 0.25 & -1.26 {\bf -1.19} -1.09 & -0.022 {\bf-0.027} -0.033 & 0.63 {\bf0.65} 0.69 \\
\hline
\end{tabular}
\caption{Parameters describing the blue tilt for the different fitting methods in different $\Mc$ models (see equations \ref{eqn:mu_tilt} and \ref{eqn:color_tilt}), in both mass-metallicity and magnitude-colour space (using the same metallicity calibration and color-metallicity relation as in CGL18).  We fit the blue tilt in each model galaxy individually and quote the 25th, 50th, and 75th percentiles of the distribution obtained from fitting all model galaxies. All fits were calculated using a minimum mass $\Mlim = 10^{5.15}\Msun$.}
  \label{tab:tilt}
\end{table*}

Observations of extragalactic GCs compiled by \citet{mieske_etal_2010} directly measure the slope of GC $(g-z)$ colour as a function of the $z$-band magnitude:
$$ \gamma_z \equiv \frac{d(g-z)}{dM_z}. $$ 
We can convert our mass-metallicity slope $\alpha$ to this equivalent observational proxy using the colour-metallicity relation adopted in CGL18:
\begin{equation}
  (g-z) = c_0 + c_1 \,\feh + c_2 \,\feh^2,
  \label{eqn:colour_met}
\end{equation}
where $c_0 = 1.513$, $c_1 = 0.481$, $c_2 = 0.051$. To convert magnitudes to cluster masses, we use the colour-dependent mass-to-light ratio of \citet{bell_etal03}:
$$
  \log_{10}M/L_z = a_{z} + b_{z}\, (g-z),
$$
where $a_z = -0.171$ and $b_z = 0.322$. Analytic transformations lead to the expression for the metallicity slope:
$$
   \alpha = \frac{d\feh}{d\log_{10}M} = \frac{-2.5\,\gamma_z}{1 - 2.5\, b_z \gamma_z} \frac{1}{c_1 + 2c_2\feh_{\rm blue}},
$$
where the average value for the blue clusters is $\feh_{\rm blue} \approx -1.5$.

Using the above conversions, we recast our fit for the metallicity as a function of cluster mass in terms of the equivalent observational quantities, $(g-z)$ and $M_z$:
\begin{equation}
   (g-z) = \gamma_z M_z + \delta_z.
   \label{eqn:color_tilt}
\end{equation}

On the right axis of \autoref{fig:tilt_slope} we convert $\alpha$ to $\gamma_z$ using the above relations. The grey shaded band shows the result found by \citet{mieske_etal_2010}: $\gamma_z = -0.029 \pm 0.0085$ for the stacked sample of all clusters with $M_z < -8.1$, corresponding to $\Mlim \approx 10^{5.15}$ for the \citet{bell_etal03} mass-to-light ratios. Using the Equal-N-Bin method, which was adopted by \citet{mieske_etal_2010}, the predicted best-fit slopes for the stacked sample of all clusters in our model are somewhat higher than their median value, but still generally consistent within the errors. 

Furthermore, comparison to any single set of observations of the blue tilt is insufficient, because observations show that the strength of the blue-tilt is not universal, but rather varies strongly between galaxy, even at fixed galaxy mass \citep{strader_etal06, cockcroft_etal_2009, mieske_etal_2010}. Indeed, several galaxies, including the Milky Way, show no blue-tilt. Therefore, it is more meaningful to compute the blue tilt slope for GC samples of individual galaxies, and then compare the distribution of slopes for a given set of galaxies.

\begin{figure}
\includegraphics[width=\columnwidth]{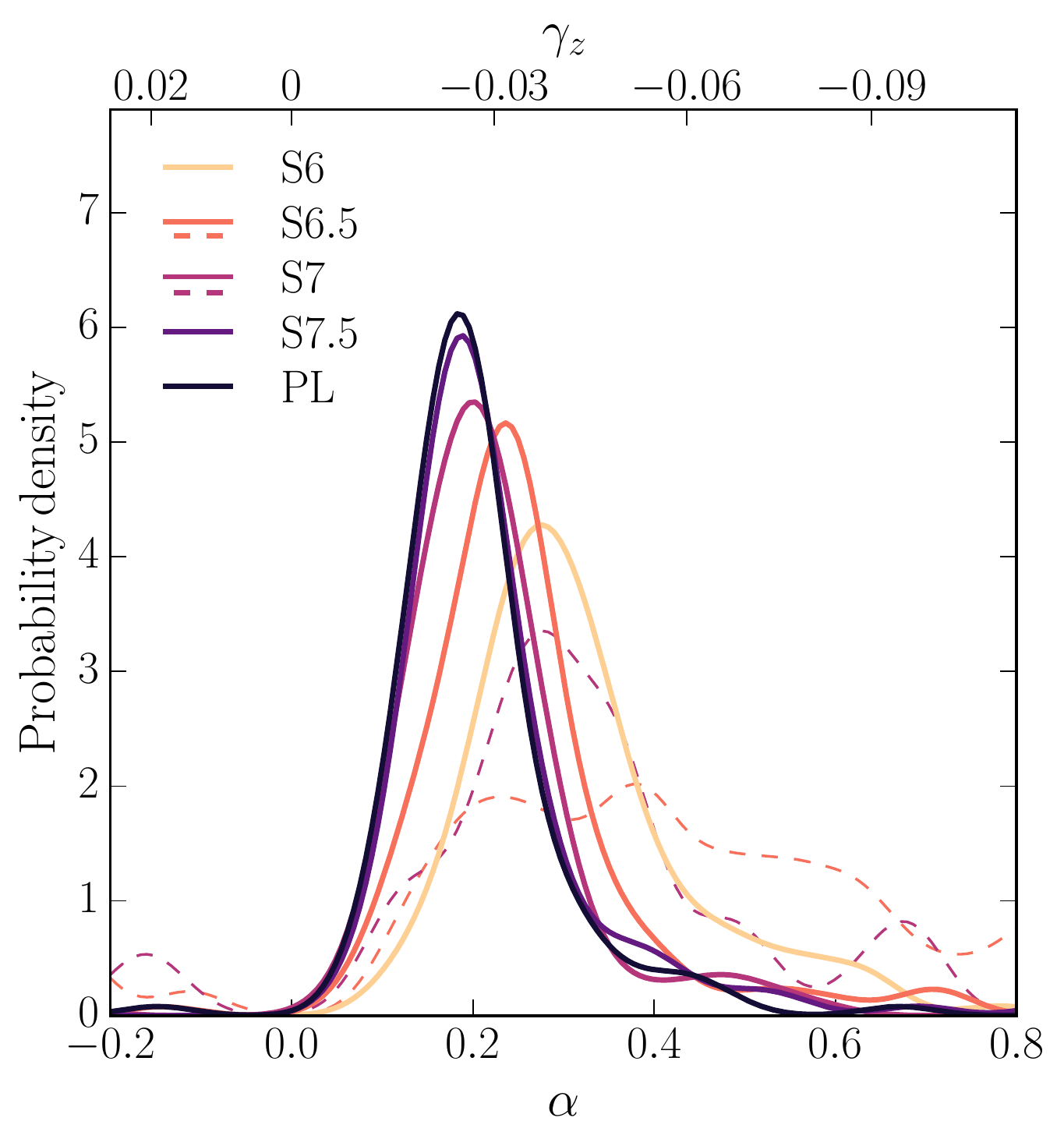}
\vspace{-5mm}
\caption{Distribution of $\alpha$ using the Single-Split method for all model galaxies with greater than 50 clusters and using a minimum mass for fitting $\Mlim = 10^{5.15} \Msun$. At lower values of $\Mc$, the distribution shifts towards higher values of $\alpha$, consistent with the expectation discussed in \autoref{sec:bluetilt}. The Equal-N-Bin method, shown in the dashed curves, gives systematically higher values of $\alpha$ than our Single-Split method.}
  \label{fig:slope_hist}
\end{figure}

For this reason we introduce a new fitting method (hereafter ``Single-Split"). 
For each galaxy, we perform a GMM fit to the metallicity distribution of \textit{all} clusters in the galaxy to determine the peak metallicities of the blue and red subpopulations. We label as blue all clusters in the region where the \rev{value of the metal-poor Gaussian} is larger than that of the metal-rich Gaussian. Finally, we fit the metallicity as a function of cluster mass for all the blue clusters above the threshold mass $\Mlim$. Thus, this method differs significantly from the Equal-N-Bin and Fixed-Bin methods because we perform only a single GMM split for each galaxy, rather than once for each cluster mass bin, as well as use the actual cluster metallicities instead of the mean value in bins for our linear fits. We impose the same lower limit on the number of clusters in a galaxy needed to perform a GMM split as we do for a single bin in the Equal-N-Bin and Fixed-Bin methods, $N_{\rm min, GMM} = 50$.

This method has several advantages over the Equal-N-Bin and Fixed-Bin methods. By performing the linear fit on all blue clusters, rather than on mean values for each bin, we can more accurately quantify the intrinsic scatter in the blue-tilt. Furthermore, this method can be reliably applied to galaxies with fewer clusters, because it requires only a single GMM split. Finally, the effects of merging of the blue and red peaks are mitigated because we perform the fit only for metallicities where blue clusters dominate. 

\begin{figure}
\includegraphics[width=\columnwidth]{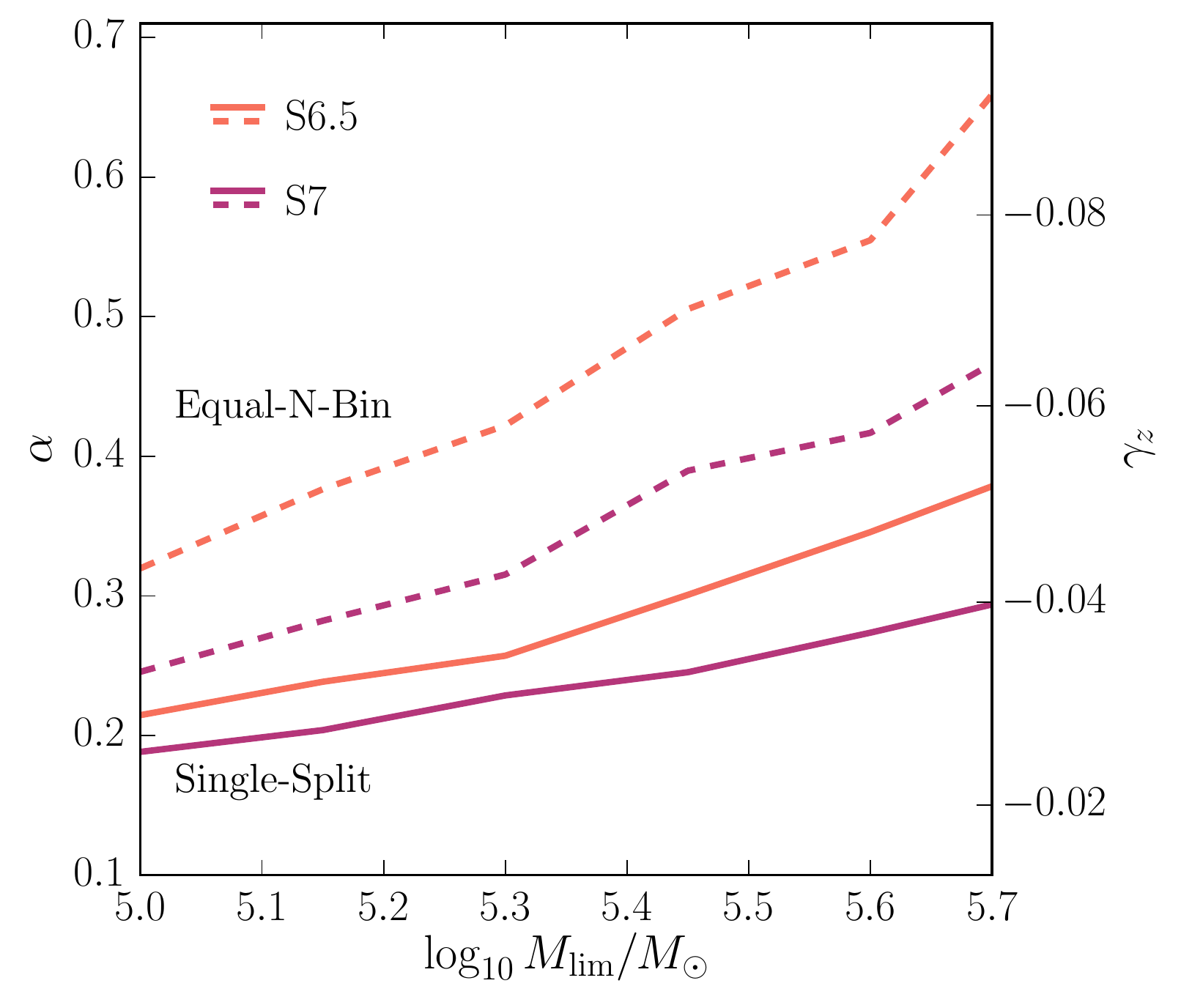}
\vspace{-5mm}
\caption{Dependence of the blue-tilt slope $\alpha$ on the minimum mass $\Mlim$ used in fitting the blue tilt. We show the median value of $\alpha$ across all galaxies. Solid lines show the result from our Single-Split method, dashed lines show the result from the Equal-N-Bin method.}
  \label{fig:mlim}
\end{figure}

\autoref{fig:slope_hist} shows the distribution of tilt slopes for all model galaxies with the Single-Split method. There is a a wide variance of the outcomes, similar to the observations. For example, in the PL model, the distribution of slopes peaks at a value of $\alpha \approx 0.18$ but also includes galaxies with $\alpha \approx 0$, which means no tilt. \rev{We note that we have also fit the Galactic GC population with the Single Split method and find it to be consistent with $\alpha \approx 0$.}

Analogously with the result for the stacked sample of all clusters (red curve in \autoref{fig:tilt_slope}), smaller values of $\Mc$ shift the peak of $\alpha$ to larger values and significantly broaden the distribution. Dashed curves in \autoref{fig:slope_hist} plot the result of the Equal-N-Bin method applied to individual galaxies, and clearly show that that method produces systematically higher values of $\alpha$ than the Single-Split, with a larger scatter.

Table~\ref{tab:tilt} lists the median values of the best-fit parameters $\alpha$ and $\beta$, and their corresponding observational counterparts $\gamma_z$ and $\delta_z$, for the three different fitting methods for our two preferred $\Mc$ models.

In the above discussion, we used a constant minimum cluster mass $\Mlim = 10^{5.15} \Msun$ in fitting the blue tilt. However, several observational studies have noted that $\alpha$ increases with increasing $\Mlim$ \citep[e.g.,][]{mieske_etal_2010}, suggesting the blue-tilt is actually non-linear. Such behaviour is a natural prediction of our model, because the formation of more massive clusters depends even more sensitively on galaxy mass, and therefore metallicity. \autoref{fig:mlim} shows that the median value of $\alpha$ indeed increases with $\Mlim$ in the S6.5 and S7 models. We find that the value of $\alpha$ scales more strongly with $\Mlim$ in lower $\Mc$ models, where the number of very massive blue clusters is smaller. We also find that this trend is stronger in the Equal-N-Bin method compared to the Single-Split method. This effect should be taken into account when comparing the results on the blue tilt in surveys or models with different limiting cluster mass.

\subsection{On the origin of blue tilt}
\label{sec:discussion}

The variation in the strength of the blue-tilt at $z=0$ reflects the assembly history of the host galaxy: if the cold gas reservoirs at high redshift were relatively low then a strong cluster mass-metallicity relation should be produced.

\begin{figure}
\includegraphics[width=\columnwidth]{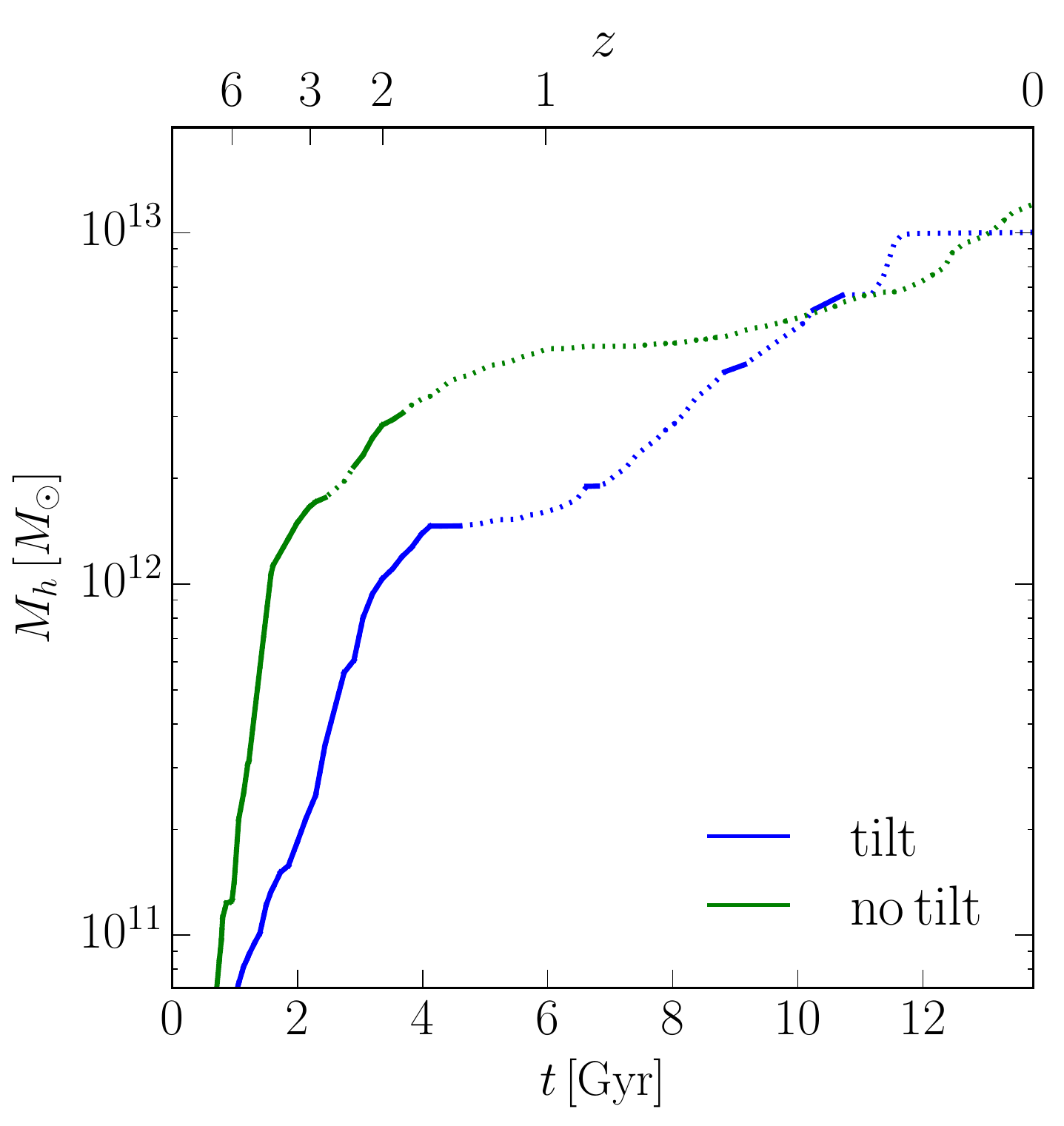}
\vspace{-5mm}
\caption{Mass growth of two halos with approximately the same mass at $z=0$, $\Mh \approx 10^{13}\Msun$. Solid regions along the curve show the periods when GCs are actively forming (i.e., the condition $R_{\rm m} > p_3$ is satisfied). The halo depicted in blue shows a blue-tilt at $z=0$, while the halo in green does not. No mass-metallicity correlation arises in the green halo because it is massive enough at early times to have a sufficient amount of cold gas for the formation of massive blue clusters.}
  \label{fig:bluetilt_tracks}
\end{figure}

To illustrate this effect, we show in \autoref{fig:bluetilt_tracks} the evolution of the main-branch of two halos, which we label ``Halo A" (green) and ``Halo B" (blue) of the same mass at $z=0$, but with very different assembly histories. At redshifts $z>1$, Halo A is overmassive relative to Halo B. In accordance with the arguments outlined above, Halo A shows zero blue-tilt, while Halo B has a typical blue-tilt: $\alpha \approx 0.2$ (using the Single-Split method). Because Halo A's cold gas reservoir -- which, by our adopted galactic scaling relations, is set by the galaxy stellar mass, in turn set by the halo mass -- was already large enough at high redshift, the CIMF could be fully sampled. This allows for the formation of massive, metal-poor clusters in Halo A. In contrast, Halo B's cold-gas reservoirs were relatively small, and thus massive clusters formed preferentially later, when Halo B was more massive and more metal-rich.

We note that for the purposes of this discussion we have compared only the clusters formed in the main branch of the halos, ignoring the contribution of satellites. In an upcoming paper, we will examine the contribution of cluster formation in satellite galaxies in more detail (N. Choksi \& O. Gnedin, in prep).

Because the metal-enhanced galaxies in which blue clusters preferentially form will be more abundant at later times, our model also predicts an ``age-tilt'' for the massive, blue clusters. We find that over the range $M = 5\times 10^5 - 5\times 10^6\Msun$, the median age of clusters decreases by 0.3~Gyr. Unfortunately, precise age measurements of extragalactic GCs are extremely difficult and have typical errors of $\approx$2 Gyr \citep[e.g.,][]{georgiev_etal12}. Improved age measurements or large samples of ages for extragalactic GCs will be required to test this model prediction.

\subsection{Comparison with other models}

Our model provides a natural explanation for the blue tilt. This removes the need for alternative models that invoke self-enrichment during the GC formation event \citep{strader_smith_2008, bailin_harris09, bailin_2018}. Of course, our results cannot rule out a possible additional contribution from self-enrichment, but they do suggest it is not needed to explain observations. 

A similar conclusion has been reached by \citet{usher_etal_2018} using the E-MOSAICS model for cluster formation, which combines analytic prescriptions for cluster formation and evolution with cosmological hydrodynamic simulations using the EAGLE model \citep{schaye_etal_2015, pfeffer_etal_2018}. They present a detailed comparison of different calibrations to convert metallicity into visual color and cluster mass to luminosity, and show that the inference of a blue tilt is robust. They find essentially the same mean slope of the blue tilt for Milky Way-sized galaxy, $\gamma_z \approx -0.03$, and a considerable scatter among different galaxy realizations.

It is interesting and reassuring that these similar conclusions are reached despite some differences in the construction of our model and that of \citet{usher_etal_2018}. The CIMF cutoff mass in their simulations is not constant but varies with local properties of the ISM using the model of \cite{reina-campos_kruijssen_2017}. The average $\Mc$ increases with cluster metallicity and host galaxy mass, and ranges between $10^6$ and $10^{7.5}\Msun$. Thus, in their model the blue tilt appears as a consequence of different cutoff masses for red and blue clusters, whereas in our model it is due to the insufficient gas supply in the host galaxies of blue clusters.

\rev{However, \citet{usher_etal_2018} find a stronger tilt in more massive galaxies and at smaller galactocentric radii, which also favor higher $\Mc$.} This trend is opposite to our results, where a stronger tilt is found for lower $\Mc$. A caveat to this comparison is that their cluster metallicity distribution does not show an obvious bimodality, so that the selection of clusters to be ``blue" is skewed towards higher metallicity. The average color of their sample of blue clusters, $(g-z)\approx 1$, corresponds to $\feh\approx -1.2$ on our metallicity scale and is instead close to the boundary separating the red and blue cluster populations in our model. 

We note that for a \textit{fixed} $\Mc$ model, we find no correlation between galaxy mass and strength of the blue-tilt, regardless of the binning method used. The lack of correlation is expected because the median halo mass in which blue clusters form is independent of the $z=0$ halo mass (see Fig.~8 of CGL18). As a result, the cold gas available for their formation is also essentially constant as the $z=0$ halo mass varies. \rev{Thus a \textit{fixed} $\Mc$ model cannot match the trend emerging in available observations that the typical blue-tilt strength increases in higher mass galaxies \citep[][]{mieske_etal_2010}. However, this does not invalidate our overall model framework. We expect the value of $\Mc$ to not be a constant, but instead to increase with galaxy mass \citep[e.g.,][]{johnson_etal_2017}, which could potentially alleviate this tension with the observations. In the present work we chose to keep $\Mc$ fixed across the entire galaxy mass range, to maintain simplicity of the model.}

\rev{It would be interesting to test these different predictions of the two models by future observations. Such a test would help in two ways. The \cite{usher_etal_2018} predictions result from the parametric cluster model coupled with simulations of galaxy formation; therefore, it is a test of its prescriptions for cluster formation and disruption as well as the physics of the underlying simulations. The predictions of our model result from the adopted galactic scaling relations; therefore, it is a test of the uncertainties of their extrapolation to high redshift. Both of these tests are important for improving our understanding of galaxy formation at high redshift. In this sense, the two models provide complementary probes.}

\section{Summary}
\label{sec:summary}

We have introduced a cutoff of the cluster initial mass function in our model of globular cluster formation and evolution. Our main results are:
\begin{enumerate}
\item Fixed cutoff masses of $\Mc = 10^{6.5}\Msun$ or $\Mc = 10^{7}\Msun$ matches many observed scaling relations, including the GC system mass-host halo mass relation, the average metallicity of the GC system-host halo mass relation, and the cluster mass functions. This range of the cutoff mass agrees with the indirect measurements of the initial mass function of GCs in the Virgo and Fornax cluster galaxies as well as several nearby galaxies.

\item Models with $\Mc < 10^{6.5}\Msun$ cannot reproduce the observed GC metallicity and mass distributions in massive early-type galaxies. Models with $\Mc > 10^7\Msun$ produce results similar to the model without a cutoff (i.e., the power-law model) and are inconsistent with the observational constraints on $\Mc$.

\item The peak of the GC formation rate density occurs about 2~Gyr earlier than that of the field star formation rate density, and corresponds to $z \approx 4-6$. 

\item Lower $\Mc$ leads to a higher total mass formed in GCs at high redshift, to compensate for the increased effect of disruption of the many small clusters.

\item The slope of the mass-metallicity relation for metal-poor clusters (blue tilt) for all $\Mc$ models is consistent with the observations within the errors, when measured using the same method: fitting peak metallicities (or colors) of the GMM mode for blue clusters in bins of cluster mass with an equal number of clusters in each bin. Using alternative methods, either with a fixed bin size or a single GMM split for all cluster masses, reveals the trend that the typical tilt strength increases with decreasing $\Mc$. 

\item The spread of the tilt slope values of GC systems for individual galaxies also increases for lower $\Mc$. We find no clear correlation between the tilt slope and galaxy mass at fixed $\Mc$.

\item In our model, the blue tilt arises because the metal-poor clusters form in relatively low-mass galaxies which lack sufficient cold gas to sample the CIMF at the highest masses. Massive blue clusters form in progressively more massive galaxies and inherit their higher metallicity. The metal-rich clusters do not exhibit such a tilt because they form in significantly more massive galaxies, which have enough cold gas to fully sample the CIMF.
\end{enumerate}

These results confirm that our simple model provides a good description of the origin of most observed scaling relations of GC systems. The introduction of the fixed cutoff of the cluster initial mass function makes the model predictions more realistic, while retaining the simplicity of only two adjustable parameters.

\section*{Acknowledgements}

We are grateful to Hui Li, Cliff Johnson, and the participants of the KITP program ``The Galaxy-Halo Connection Across Cosmic Time" in June 2017 for useful discussions, as well as Diederik Kruijssen for a constructive referee report which improved this work. Research at KITP is supported in part by the National Science Foundation through grant PHY17-48958. We thank the Illustris team for releasing their halo catalogs and Benedikt Diemer for making the \textsc{colossus} code publicly available. NC thanks Goni Halevi for helpful comments and support throughout the preparation of this work. This work was supported in part by the National Science Foundation through grant 1412144.




\bibliographystyle{mnras}
\bibliography{GC/gc_oleg,GC/gc_nick}



\appendix

\section{Effect of the CIMF sampling methods on the blue tilt}
\label{sec:appendix_sampling}

In \autoref{sec:sampling}, we discussed our application of the ``optimal sampling'' method for drawing cluster masses from the CIMF. This method sets the maximum cluster mass deterministically, via \autoref{eqn:constraint2}. Since the blue tilt is affected by massive clusters, here we test how our conclusions are sensitive to the method used for drawing cluster masses. We test two alternative sampling methods. The first is pure random sampling without drawing $\Mmax$ first, while the second does an effective Poisson sampling. 

\begin{figure}
\includegraphics[width=\columnwidth]{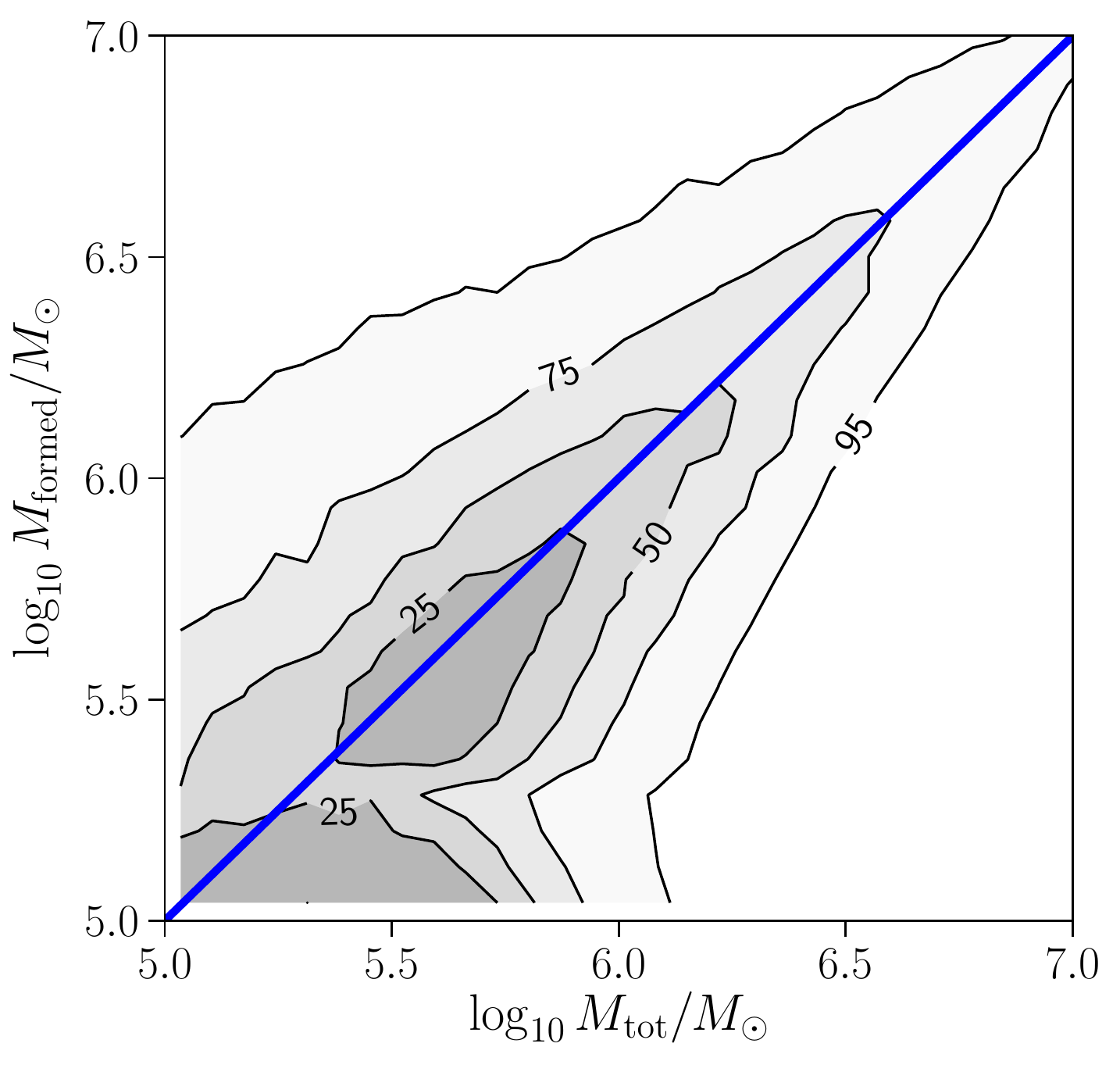}
\vspace{-5mm}
\caption{The total mass actually formed in clusters from sampling the CIMF (i.e., sum of individual cluster masses) using the Poisson method vs. the total mass $\Mtot$ that was supposed to be formed, given by \autoref{eqn:mgc}. The black line shows the one-to-one relation $M_{\rm formed} = M_{\rm tot}$. All cluster formation events in one realization of the model are shown, and the mass range is limited to the typical values for $M_{\rm tot}$ during events in which blue clusters form. Contours labels represent the percent of all points enclosed.}
  \label{fig:poisson_mass}
\end{figure}

In the Poisson method, described in \cite{sormani_etal_2017}, we first calculate the expected mean cluster mass and number of clusters:
\begin{align}
  \bar{N} &\equiv \frac{\Mtot}{\bar{M}} \nonumber\\
  \bar{M} &\equiv {\int_{\Mmin}^{\Mu} M\frac{dN}{dM} \bigg/ \int_{\Mmin}^{\Mu} \frac{dN}{dM}} dM \nonumber\\
    &= M_c \frac{\Gamma(2-\beta, \Mmin/\Mc)-\Gamma(2-\beta, \Mu/\Mc)}{\Gamma(1-\beta, \Mmin/\Mc)-\Gamma(1-\beta, \Mu/\Mc)}. \nonumber
\end{align}
The last line applies for a Schechter CIMF as parameterized in \autoref{eqn:cimf}. We consider the effects of different values of the upper limit of integration $\Mu$ below. We then sample $N_{\rm samp}$ clusters from the CIMF, where $N_{\rm samp}$ is drawn from a Poisson distribution with mean $\bar{N}$. Hence, this method does not enforce strict mass conservation: the sum of cluster masses formed in a single event can deviate substantially from $\Mtot$. This is demonstrated by the large scatter around the one-to-one line in \autoref{fig:poisson_mass}. 

This Poisson method is suitable for sampling the stellar IMF in GMC-scale simulations of star formation using sink particles. These sinks grow continuously by small increments of mass, each of which may not fully sample the IMF. In contrast, our model forms a complete population of clusters in a given merger episode, which lasts much longer than a timescale for forming individual clusters. Therefore, although we have tested this method here for completeness, we believe that in the context of our model, it is less appropriate than optimal or random sampling of the CIMF.

In all three cases, we draw clusters by numerically inverting the cumulative distribution function given by:
\begin{equation}
\frac{N(<M)}{N(<\Mu)} = \frac{\Gamma(1 - \beta, {\Mmin}/{\Mc}) - \Gamma(1-\beta, {M}/{\Mc})}{\Gamma(1-\beta, {\Mmin}/{\Mc}) - \Gamma(1-\beta, {\Mu}/{\Mc})}.
\label{eqn:cumulative}
\end{equation}
For optimal sampling, $\Mu = \Mmax(\Mtot)$. For Poisson and random sampling, we test two alternate methods. The first method -- hereafter ``Poisson-Fix'' and ``Random-Fix'' -- adopts a fixed value of $\Mu = 10^8 \Msun$. In the second method -- hereafter ``Poisson-$\Mmax$'' and ``Random-$\Mmax$'' -- we adopt a variable value of $\Mu = \Mmax$, where $\Mmax(\Mtot)$ is set by the same constraint (\autorefp{eqn:constraint2}) as in optimal sampling.

\begin{figure}
\includegraphics[width=\columnwidth]{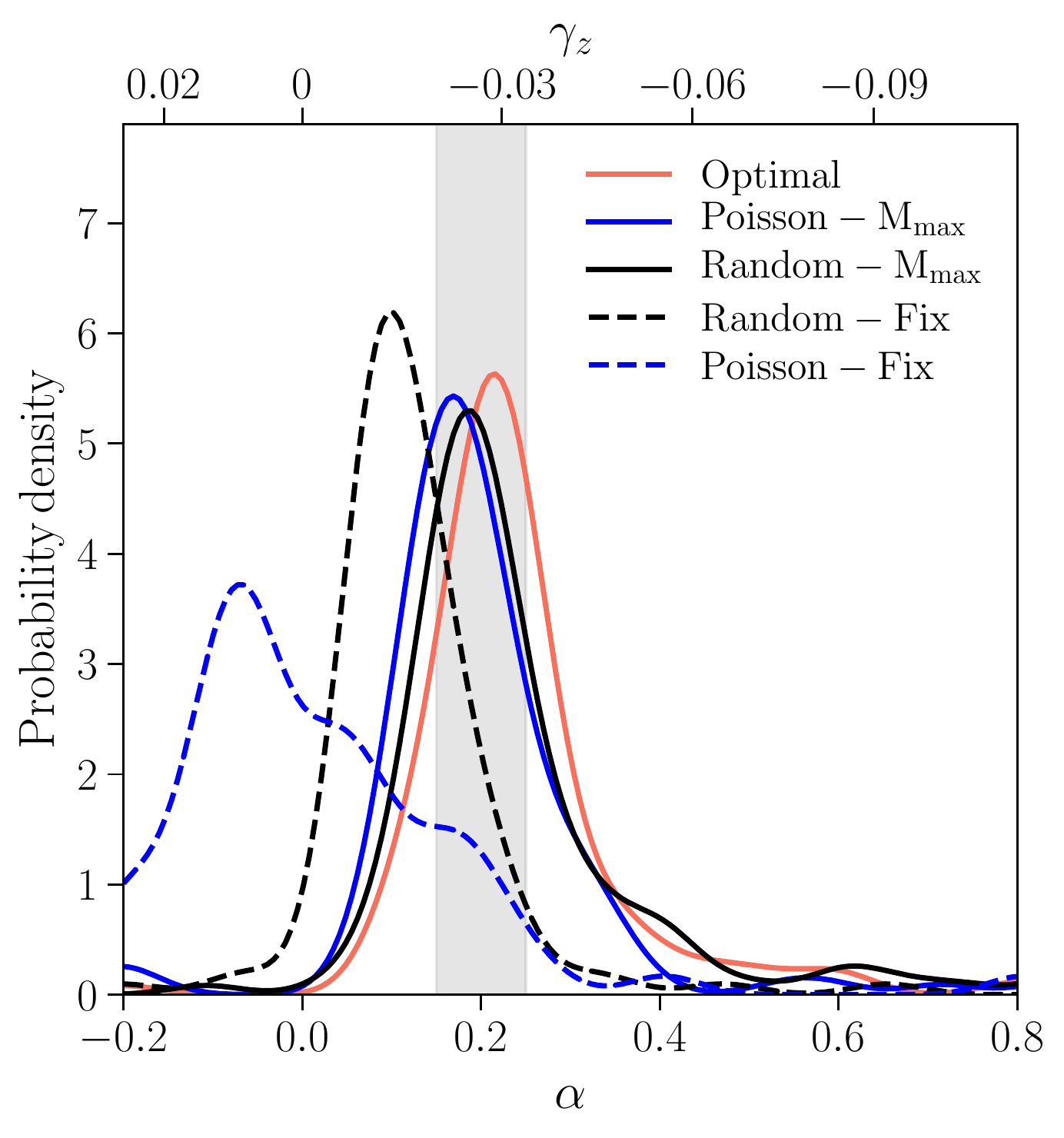}
\vspace{-5mm}
\caption{Comparison of the distribution of blue tilt slopes for all model galaxies for different sampling methods (in all cases, $\Mc = 10^{6.5} \Msun$ and $M_{\rm lim} = 10^{5.15} \Msun$). The grey shaded region highlights the $\pm 1-\sigma$ range for a \textit{single fit} of the blue tilt to the stacked sample of Virgo and Fornax clusters from \protect \cite{mieske_etal_2010}.}
  \label{fig:samp_comp}
\end{figure}

These different sampling methods lead to different predictions for the strength of the blue tilt, as shown in \autoref{fig:samp_comp}. Sampling with variable $\Mmax$ leads to slightly weaker blue tilts ($\alpha \approx 0.20$) than predicted by our fiducial optimal sampling, with only minor differences between the ``Poisson-$\Mmax$'' and ``Random-$\Mmax$'' sampling methods. On the other hand, adopting a fixed value of $\Mu = 10^8 \Msun$ leads to drastic differences relative to the variable $\Mmax$ methods. In the ``Random-Fix'' method, the center of the distribution of blue tilt slopes shifts to $\alpha \approx 0.12$, while in the ``Poisson-Fix'' method the distribution is broad and centered on zero. We discuss below the reasons for these effects.

In the limit $\Mmax \gg \Mc$, all sampling methods should produce nearly indistinguishable results. However, this is not the regime in which blue clusters form, because $\Mmax$ depends on $\Mtot$ (which in turn scales with host galaxy mass; see \autorefp{eqn:mgc}). For typical galaxies in which blue clusters form, solving \autoref{eqn:constraint3} for $\Mmax$ yields $\Mmax \lesssim \Mc$. Therefore, the imposition of this variable $\Mmax$ has a significant impact on the high-mass clusters that set the blue tilt. The quantitative impact of this change on CIMF sampling can be seen directly in \autoref{eqn:cumulative}. It leads to a lower average cluster mass because the mass function is truncated at $\Mmax$. In fact, it is the imposition of this variable $\Mmax$ which dominates the blue tilt effect seen in our model, as demonstrated by the nearly identical blue tilts for the Optimal, Random-$\Mmax$, and Poisson-$\Mmax$ models, in contrast to the Random-Fix and Poisson-Fix models.

However, even if we take a constant $\Mu = 10^8 \Msun$, random sampling (``Random-Fix'') still produces a (weak) blue tilt. This is because we enforce mass conservation when performing random sampling. As a result, in very low-mass (and therefore low-metallicity) galaxies, the formation of massive blue clusters with $M \sim \Mtot$ is still suppressed. However, the predicted blue tilt is weaker than in methods with variable $\Mmax$, because taking $\Mu = 10^8 \Msun$ increases \textit{the probability of drawing} more massive blue clusters.

In contrast to the weakening of the blue tilt in the Random-Fix method, Poisson sampling with a fixed $\Mu = 10^8 \Msun$ leads to no blue tilt. This lack of correlation is expected because Poisson sampling does not enforce mass conservation, and massive blue clusters are allowed to form regardless of their host galaxy's mass.

One caveat to the above analysis is that, because the Poisson and random sampling predict more massive clusters, which experience relatively less tidal mass loss (equation~\ref{eqn:ttid}), for the same CIMF normalization these alternate sampling methods predict more surviving clusters relative to optimal sampling. To test the magnitude of this effect, we tried re-optimizing the various sampling methods for $\Mc = 10^{6.5} \Msun$ by allowing $p_2$ (which sets the CIMF normalization) to vary. For simplicity, we kept $p_3$ fixed at the value obtained for optimal sampling. In all four cases we found only slight decreases ($p_2 \approx 12-13$) relative to the best-fit value for optimal sampling, $p_2 = 13.5$. The resulting blue tilt slopes increase by 11\% and 5\%, for the Random-$\Mmax$ and Poisson-$\Mmax$ methods. In the Random-Fix method the blue tilt increases by 8\%, while even after re-optimization the Poisson-Fix method yields no blue tilt, consistent with the arguments outlined above. Aside from the blue tilt, all sampling methods robustly reproduce all other $z=0$ GC system scaling relations. 

Which of the CIMF sampling methods is best? Currently, there is limited observational evidence in support of any one method. Overall, our results demonstrate that so long as the formation of blue clusters depends on the host galaxy mass, a blue tilt will arise. This is the case in all of the sampling methods discussed in this appendix, except for the Poisson-Fix method. The dependence on host galaxy mass comes from either imposing a variable maximum cluster mass that depends on the gas reservoir mass or simply performing gas reservoir mass-constrained sampling.


\bsp	
\label{lastpage}
\end{document}